\documentclass[trsc,nonblindrev]{informs3} % current default for manuscript submission

\OneAndAHalfSpacedXI % current default line spacing
\usepackage[utf8]{inputenc}
\usepackage[english]{babel}
\usepackage{mathtools}
\usepackage{bbm}

\usepackage[toc,page]{appendix}

    \usepackage{natbib}
     \bibpunct[, ]{(}{)}{,}{a}{}{,}%
     \def\bibfont{\small}%
     \usepackage{dsfont}
    \usepackage{amsmath}
     \usepackage{amssymb}
     
     \let\oldemptyset\emptyset
    \let\emptyset\varnothing
    
    \usepackage{enumitem}
\TheoremsNumberedThrough     % Preferred (Theorem 1, Lemma 1, Theorem 2)
\EquationsNumberedThrough    % Default: (1), (2), ...
\renewcommand{\theARTICLETOP}{}
\renewcommand{\theMANUSCRIPTNO}{}

\begin{document}
%%%%%%%%%%%%%%%%

%\setcounter{page}{1}
%\DOI{10.1287/xxxx.0000.0000}%


\RUNTITLE{Risk-Averse Equilibrium for Autonomous Vehicles in Stochastic Congestion Games}

\RUNAUTHOR{Yekkehkhany and Nagi}

% \TITLE{Bundling Information Goods of Decreasing Value}
\TITLE{Risk-Averse Equilibrium for Autonomous Vehicles in Stochastic Congestion Games}

\ARTICLEAUTHORS{%
\AUTHOR{Ali Yekkehkhany}
\AFF{Electrical and Computer Engineering, University of Illinois at Urbana-Champaign, \EMAIL{yekkehk2@illinois.edu}%, \URL{}
}
\AUTHOR{Rakesh Nagi}
\AFF{Industrial and Enterprise Systems Engineering, University of Illinois at Urbana-Champaign, \EMAIL{nagi@illinois.edu}%, \URL{}
}
} % end of the block

\ABSTRACT{%
The fast-growing market of autonomous vehicles, unmanned aerial vehicles, and fleets in general necessitates the design of smart and automatic navigation systems considering the stochastic latency along different paths in the traffic network. The longstanding shortest path problem in a deterministic network, whose counterpart in a congestion game setting is Wardrop equilibrium, has been studied extensively, but it is well known that finding the notion of an optimal path is challenging in a traffic network with stochastic arc delays. In this work, we propose three classes of risk-averse equilibria for an atomic stochastic congestion game in its general form where the arc delay distributions are load dependent and not necessarily independent of each other. The three classes are risk-averse equilibrium (RAE), mean-variance equilibrium (MVE), and conditional value at risk level $\alpha$ equilibrium (CVaR$_\alpha$E) whose notions of risk-averse best responses are based on maximizing the probability of taking the shortest path, minimizing a linear combination of mean and variance of path delay, and minimizing the expected delay at a specified risky quantile of the delay distributions, respectively. We prove that for any finite stochastic atomic congestion game, the risk-averse, mean-variance, and CVaR$_\alpha$ equilibria exist. We show that for risk-averse travelers, the Braess paradox may not occur to the extent presented originally since players do not necessarily travel along the shortest path in expectation, but they take the uncertainty of travel time into consideration as well. We show through some examples that the price of anarchy can be improved when players are risk-averse and travel according to one of the three classes of risk-averse equilibria rather than the Wardrop equilibrium.
}%

%\KEYWORDS{deterministic inventory theory; infinite linear programming duality; 
%  existence of optimal policies; semi-Markov decision process; cyclic schedule}

%%%%%%%%%\HISTORY{}

\maketitle
%%%%%%%%%%%%%%%%%%%%%%%%%%%%%%%%%%%%%%%%%%%%%%%%%%%%%%%%%%%%%%%%%%%%%%

% Samples of sectioning (and labeling) in TRSC
% NOTE: (1) \section and \subsection do NOT end with a period
%       (2) \subsubsection and lower need end punctuation
%       (3) capitalization is as shown (title style).
%
%\section{Introduction.}\label{intro} %%1.
%\subsection{Duality and the Classical EOQ Problem.}\label{class-EOQ} %% 1.1.
%\subsection{Outline.}\label{outline1} %% 1.2.
%\subsubsection{Cyclic Schedules for the General Deterministic SMDP.}
%  \label{cyclic-schedules} %% 1.2.1
%\section{Problem Description.}\label{problemdescription} %% 2.

% Text of your paper here
\section{Introduction}
\label{introduction}

The intelligent transportation systems are growing faster than ever with the speedy emergence of autonomous vehicles, unmanned aerial vehicles, Amazon delivery robots, Uber/Lyft self-driving cars, and such.
One of the principal components of such systems is the navigation system whose goal is to  provide travelers with fast and reliable paths from their sources to destinations.
In a fleet of vehicles, an equilibrium is achieved when no travelers have any incentives in a certain sense to change routes unilaterally.
In the classical Wardrop equilibrium \citep{wardrop1952correspondence, wardrop1952road}, travelers have incentives to change routes if they have an alternative route that has lower expected travel time.
%%%%%%%%%%%\citep{} will generate parantheses
In other words, the optimality metric is based on minimizing the expected travel time in the Wardrop equilibrium.
In the context of transportation though, collisions, weather conditions, road works, traffic signals, and varying traffic conditions can cause deviations in travel times \citep{ordonez2010wardrop}.
As a result, the path with the minimum expected travel time may not be reliable due to its high variability.
Similarly, in the context of telecommunication networks, noise, signal degradation, interference, re-transmission, and malfunctioning equipment can cause variability in transmission time from source to destination \citep{ordonez2010wardrop}.
The empirical works by 
%Abdel-Aty et al.
\cite{abdel1995investigating}, 
%Kazimi et al. 
\cite{kazimi2000van}, 
%Lam \citep{lam2000effect}, 
%Lam and Small 
\cite{lam2001value}, and 
%Small 
\cite{small1999valuation} also support the fact that taking travel time uncertainty into account is indeed an essential criterion in navigation systems.

As mentioned above, minimizing the expected travel time is inadequate in scenarios involving risk due to variability of travel times.
In order to address this issue, we study a richer class of congestion games called stochastic congestion games in an atomic setting, where the travel times along different arcs of the network are random variables that are not necessarily independent of each other.
In this framework, we introduce probability statements regarding the risk-averse best response of a traveler given the choice of the rest of travelers in the network.
We propose three classes of risk-averse equilibria for stochastic congestion games: risk-averse equilibrium (RAE), mean-variance equilibrium (MVE), and conditional value at risk level $\alpha$ equilibrium (CVaR$_\alpha$E), whose notions of risk-averse best responses are based on maximizing the probability of traveling along the shortest path (also known as Risk-Averse Best Action Decision with Incomplete Information (R-ABADI)), minimizing a linear combination of mean and variance of path delay, and minimizing the expected delay at a specified risky quantile of the delay distributions, respectively.
We prove that the risk-averse, mean-variance, and CVaR$_\alpha$ equilibria exist for any finite stochastic atomic congestion game.
Note that two equilibria similar to the mean-variance and CVaR equilibria exist in the literature and are discussed in the related work section, but the probability distributions of travel times are load independent or link delays are considered to be independent in the literature, which is not the case in this article.
It is noteworthy that most studies on stochastic congestion games make use of simplifying assumptions such as considering the arc delay distributions to be independent of their loads or adding independent and identically distributed errors to nominal delays of arcs neglecting their differences.
In the Braess paradox \citep{braess1968paradoxon, murchland1970braess}, which is known to be a counterintuitive example rather than a paradox, the risk-neutral/selfish travelers select the shortest path in expected travel time, which maximizes the social delay/cost incurred by the whole society.
Although the focus of this article is not on deriving bounds on price of anarchy, we study the Braess paradox in a stochastic setting under the three proposed risk-averse equilibria and show that the risk-averse behavior of travelers results in improving the social delay/cost incurred by the society; and as a result, the price of anarchy is improved if travelers are risk-averse.
As the result, the Braess paradox may not occur to the extent presented originally if travelers are risk-averse.
Furthermore, we study the Pigou network \citep{pigou1920economics} in a stochastic setting and observe that the price of anarchy is also improved if travelers are risk-averse in the senses discussed above.
Note that the Pigou networks are prevalent in traffic/telecommunication networks.
Hence, providing travelers with risk-averse navigation can decrease the social delay/cost in the real world applications.

The article is structured in the following way.
The related work is discussed in Section \ref{related_work_RAE_Congestion_Games}.
The stochastic congestion game is formally defined in Section \ref{problem_statement_congestion_game}.
The three proposed classes of equilibria, i.e. risk-averse, mean-variance, and CVaR$_\alpha$ equilibria, are presented in Section \ref{section_risk_averse_equilibrium_congestion_game} and their existences in any finite stochastic congestion game are proven; detailed proofs can be found in the Appendix.
Numerical results including the study of the Pigou and Braess networks as well as notes for practitioners are provided in Section \ref{numerical_results_congestion_game}.
Finally, conclusions and discussion of opportunities for future work are provided in Section \ref{conclusion_future_congestion_game}.

\section{Related Work}
\label{related_work_RAE_Congestion_Games}

In this section, the literature on navigation for both deterministic and stochastic networks is presented first, then the literature on deterministic and stochastic congestion games is discussed in details.
The main focus of the literature review is to motivate the necessity of risk-averse algorithms for navigation and congestion games in a stochastic setting.

The problem of finding the shortest path in a transportation/telecommunication traffic network is one of the main parts of the in-vehicle navigation systems.
This problem has been studied well in deterministic networks resulting in many efficient algorithms, e.g., the algorithms developed by 
%Bellman \citep{bellman1958routing},
%Dijkstra 
\cite{dijkstra1959note}, and 
%Dreyfus 
\cite{dreyfus1969appraisal}; also see \citep{schrijver2012history, fu2006heuristic, dial1969algorithm, tarjan1983data, lawler1976combinatorial, pierce1975bibliography, orda1990shortest, kaufman1993fastest, hall1993time, chabini1997new}, and \citep{hosseini2017mobile}.
Although finding the shortest path problem is well understood in deterministic networks, the definition of an optimal path and how to identify such a path is more challenging in the stochastic version of the problem.
There have been multiple approaches to define the optimal path in stochastic networks as summarized below.
The least expected travel time is studied by 
%Loui 
\cite{loui1983optimal} and is equivalent to the deterministic case from a computational point of view.
The path with the least expected time may be sub-optimal for risk-averse travelers due to its high variability and uncertainty; as the result, the probability distributions of link travel times need to be considered explicitly to find the most reliable path.
In this manner, 
%Frank 
\cite{frank1969shortest} proposed the optimal path to be the one that maximizes the probability of realizing a travel time that less than a threshold, 
%Sigal et al. 
\cite{sigal1980stochastic} proposed the optimal path to be the one that maximizes the probability of realizing the shortest time, and 
%Chen and Ji 
\cite{chen2005path} proposed the optimal path to be the one with minimum travel time budget required to meet a travel time reliability constraint. For more variants of the mentioned algorithms, refer to \citep{nie2006arriving, nie2009shortest, zeng2015application, xing2011finding, howard2012dynamic, hall1986fastest, fu1998expected, waller2002online, miller2000least, mirchandani1986routes, mirchandani1976shortest, murthy1996relaxation, fan2003optimal, xiao2013adaptive, bell2009hyperstar, chen2010risk}, and \citep{lo2006degradable}.

%Despite the fact that networks with stochastic arc delays has been studied in route selections, they have not much attention is given to been studied in a game setting.

In the context of route selection in a fleet of vehicles, a game emerges between all travelers where the action of each traveler affects the travel time of the other travelers, which creates a competitive situation forcing travelers to strategize their decisions.
In a deterministic network, the mentioned game is formalized by 
%Wardrop and Whitehead
\cite{wardrop1952correspondence}, 
%von Neumann and Morgenstern
\cite{neumann1928theorie, von1947theory}, and 
%Nash et al. 
\cite{nash1950equilibrium}.
However, it is not realistic to consider the link delays to be known prior to making a decision due to external factors that make the travel times uncertain.
In order to put this in perspective, several approaches have been adopted by researchers to capture the stochastic behavior of the traffic networks. For example, 
%Harsanyi 
\cite{harsanyi1967games, harsanyi1968games} proposed Bayesian games that consider the incomplete information of payoffs, 
%Ord{\'o}{\~n}ez and Stier-Moses
\cite{ordonez2010wardrop} modeled the risk-averse behavior of travelers by padding the expected travel time along paths with a safety margin,
%Watling 
\cite{watling2006user} proposed an equilibrium based on the optimality measure of minimizing the probability of being late or maximizing the probability of being on time,
%Szeto et al. 
\cite{szeto2006risk} associated a cost with the travel time uncertainty based on travelers' risk-averse behavior,
%Chen and Zhou 
\cite{chen2010alpha} proposed an equilibrium based on the optimality measure of minimizing the conditional expectation of travel time beyond a travel time budget,
and 
%Bell and Cassir 
\cite{bell2002risk} proposed to play out all possible scenarios before making a choice.
For more details in the context of traffic networks, we refer readers to \citep{aashtiani1981equilibria, aghassi2006robust, altman2006survey, hayashi2005robust, mirchandani1987generalized, nie2011multi, connors2009network, schmocker2009game, fonzone2012link, angelidakis2013stochastic, nikolova2011stochastic, nikolova2015burden}, and \citep{correa2019network}.

%Despite the fact that it has been known for quite a long time that route latency is a random variable than a deterministic number \citep{frank1969shortest}, there has not been much efforts to formalize an equilibrium for stochastic congestion games that takes the probability distributions into account.
%traverse

\section{Problem Statement}
\label{problem_statement_congestion_game}
Consider a directed graph (network) $G = (\mathcal{N}, \mathcal{E})$ with a node set $\mathcal{N} = [N] \coloneqq \{1, 2, \dots, N\}$ and directed link (edge) set $\mathcal{E}$ with cardinality $|\mathcal{E}|$, where the pair $(i, j) \in \mathcal{E}$ indicates a directed link from node $i \in \mathcal{N}$ to node $j \in \mathcal{N}$ in the directed graph.
Denote the set of source-destination (SD) pairs with $\mathcal{K} \subseteq \mathcal{N} \times \mathcal{N}$, where for the SD pair $k = (s_k, d_k) \in \mathcal{K}$, $s_k \neq d_k$, the set of simple directed paths from $s_k$ to $d_k$ in $G$ is denoted by $\mathcal{P}_k$, and let $n_k$ be the number of players (travelers, vehicles, or data packages) associated with source-destination $k$.
Let $\mathcal{P} \coloneqq \cup_{k \in \mathcal{K}} \mathcal{P}_k$ be the set of all paths.
A feasible assignment $\boldsymbol{m} \coloneqq \{ m^p : p \in \mathcal{P} \}$ allocates a non-negative number of players to every path $p \in \mathcal{P}$ such that $\sum_{p \in \mathcal{P}_k} m^p = n_k$ for all $k \in \mathcal{K}$.
As a result, the number of players along link $e \in \mathcal{E}$ denoted by $m_e$ is given by $m_e = \sum_{ \{ p \in \mathcal{P}: e \in p \} } m^p$.

The latency (delay or travel time) along link $e$ is load-dependent which is denoted by the non-negative continuous random variable $L_e(m_e)$ with marginal probability density function (pdf) $f_e(x | m_e)$ and mean $l_e(m_e)$.
Note that the number of players along an edge is determined by an assignment $\boldsymbol{m}$, so $L_e(\boldsymbol{m})$, $f_e(x | \boldsymbol{m})$, and $l_e(\boldsymbol{m})$ can be used instead of $L_e(m_e)$, $f_e(x | m_e)$, and $l_e(m_e)$, respectively.
Furthermore, the latency along links of the graph can be dependent, in which case, the joint pdf of latency over all links is denoted by $f_{e_1, e_2, \dots, e_{|\mathcal{E}|}}(x_1, x_2, \dots, x_{|\mathcal{E}|} | m_1, m_2, \dots, m_{|\mathcal{E}|})$, which can be denoted as $f_\mathcal{E}(x_1, x_2, \dots, x_{|\mathcal{E}|} | \boldsymbol{m})$.
Given the link latency defined above, the nominal latency of player $i$ along path $p_i \in \mathcal{P}$ under a given assignment $\boldsymbol{m}$ is simply $L^i(\boldsymbol{m}) \coloneqq \sum_{e \in p_i} L_e(\boldsymbol{m})$ with pdf $f^i(x | \boldsymbol{m}) = \partial \left ( \int\int\dots\int_{ \{ \sum_{e \in p_i} x_e \leq x \} } \ f_\mathcal{E}(x_1, x_2, \dots, x_{|\mathcal{E}|} | \boldsymbol{m}) \ dx_1 dx_2 \dots dx_{|\mathcal{E}|} \right ) \Big \slash \partial x$ and mean $l^i(\boldsymbol{m}) = \sum_{e \in p_i} l_e(\boldsymbol{m})$.

The stochastic congestion game consists of $n \coloneqq \sum_{k \in \mathcal{K}} n_k$ players (travelers), where player $i \in [n] \coloneqq \{ 1, 2, \dots, n \}$ is associated with the corresponding source-destination pair $k(i) \in \mathcal{K}$.
As a result, $\mathcal{P}_{k(i)}$ is the set of possible pure strategies (actions or paths) for player $i$.
The pure strategy profile of all $n$ players is denoted by $\boldsymbol{p} \coloneqq (p_1, p_2, \dots, p_n)$, where $p_i \in \mathcal{P}_{k(i)}$, that fully specifies all actions in the game.
The set of all pure strategy profiles is the Cartesian product of pure strategy sets of all players which is denoted by $\boldsymbol{\mathcal{P}} \coloneqq \mathcal{P}_{k(1)} \times \mathcal{P}_{k(2)} \dots \times \mathcal{P}_{k(n)}$.
Let $\boldsymbol{p}_{-i} \coloneqq (p_1, p_2, \dots, p_{i - 1}, p_{i + 1}, \dots, p_n)$ be the pure strategies of all players except player $i$, so $\boldsymbol{p} = (p_i, \boldsymbol{p}_{-i})$.
Given the pure strategy profile $\boldsymbol{p}$, the number of players on a path $p \in \mathcal{P}$ is given by $m^p = \sum_{i = 1}^n \mathbbm{1} \{ p_i = p \}$, and the number of players on a link $e \in \mathcal{E}$ is given by $m_e = \sum_{ \{ p \in \mathcal{P}: e \in p \} } \sum_{i = 1}^n \mathbbm{1} \{ p_i = p \}$.
Let $\boldsymbol{m}(\boldsymbol{p})$ show the number of players on all paths which is fully determined by the pure strategy $\boldsymbol{p}$.
As a result, given the pure strategy profile $\boldsymbol{p} = (p_i, \boldsymbol{p}_{-i})$, the latency of player $i$ by choosing the path $p_i$ is the random variable $L^{i} (\boldsymbol{m}(\boldsymbol{p})) = \sum_{e \in p_i} L_e(\boldsymbol{m}(\boldsymbol{p}))$ with pdf $f^{i}(x | \boldsymbol{m}(\boldsymbol{p}))$ and mean $l^{i}(\boldsymbol{m}(\boldsymbol{p})) = \sum_{e \in p_i} l_e(\boldsymbol{m}(\boldsymbol{p}))$.
For simplicity, instead of using $L^{i} (\boldsymbol{m}(\boldsymbol{p}))$, $f^{i}(x | \boldsymbol{m}(\boldsymbol{p}))$, and $l^{i}(\boldsymbol{m}(\boldsymbol{p}))$, we use $L^{i} (\boldsymbol{p})$, $f^{i}(x | \boldsymbol{p})$, and $l^{i}(\boldsymbol{p})$, respectively.

The mixed strategy of player $i$ is denoted by $\sigma_i \in \Sigma_i$, where $\Sigma_i$ is the set of all probability distributions over the set of pure strategies $\mathcal{P}_{k(i)}$, and $\sigma_i(p)$ is the probability that player $i$ selects path $p$.
The mixed strategy profile of all $n$ players is denoted by $\boldsymbol{\sigma} \coloneqq (\sigma_1, \sigma_2, \dots, \sigma_n)$, where $\sigma_i \in \Sigma_i$.
The set of all mixed strategy profiles is the Cartesian product of mixed strategy sets of all players which is denoted by $\boldsymbol{\Sigma} \coloneqq \Sigma_1 \times \Sigma_2 \dots \times \Sigma_n$.
Let $\boldsymbol{\sigma}_{-i} \coloneqq (\sigma_1, \sigma_2, \dots, \sigma_{i - 1}, \sigma_{i + 1}, \dots, \sigma_n)$ be the mixed strategies of all players except player $i$, so $\boldsymbol{\sigma} = (\sigma_i, \boldsymbol{\sigma}_{-i})$.
The latency of player $i$ by selecting path $p_i$ when the other $[n] \setminus i$ players select paths according to a mixed strategy $\boldsymbol{\sigma}_{-i}$ is denoted by the random variable $\overline{L}^{i}(p_i, \boldsymbol{\sigma}_{-i})$ that has the following pdf using the law of total probability:
\begin{equation}
\label{dist_mixed}
\bar{f}^{i}(x | (p_i, \boldsymbol{\sigma}_{-i})) = \sum_{\boldsymbol{p}_{-i} \in \boldsymbol{\mathcal{P}}_{-i}} \Big ( f^{i}(x | (p_i, \boldsymbol{p}_{-i})) \cdot \boldsymbol{\sigma}(\boldsymbol{p}_{-i}) \Big ),
\end{equation}
where $\boldsymbol{\sigma}(\boldsymbol{p}_{-i}) =  \prod_{j \in [n] \setminus i} \sigma_j (p_j)$ and $p_j$ is the corresponding strategy of player $j$ in $\boldsymbol{p}_{-i}$, and the mean of the random variable is given as
\begin{equation}
\label{eq_mean_mixed}
    \overline{l}^{i}(p_i, \boldsymbol{\sigma}_{-i}) \coloneqq \mathbbm{E}[\overline{L}^{i}(p_i, \boldsymbol{\sigma}_{-i})] = \sum_{\boldsymbol{p}_{-i} \in \boldsymbol{\mathcal{P}}_{-i}} \Big ( l^{i}(p_i, \boldsymbol{p}_{-i}) \cdot \boldsymbol{\sigma}(\boldsymbol{p}_{-i}) \Big ).
\end{equation}

The expected average delay (latency) incurred by the $n$ players in the stochastic congestion game under the pure strategy profile $\boldsymbol{p}$, also known as the \textit{social cost} or \textit{social delay} in this context, is denoted by $D(\boldsymbol{p}) \coloneqq \frac{1}{n} \sum_{i = 1}^n l^{i}(\boldsymbol{p})$.
The social delay under the mixed strategy $\boldsymbol{\sigma}$ is $D(\boldsymbol{\sigma}) \coloneqq \frac{1}{n} \sum_{\boldsymbol{p} \in \boldsymbol{\mathcal{P}}} \sum_{i = 1}^n \boldsymbol{\sigma}(\boldsymbol{p}) \cdot l^{i}(\boldsymbol{p})$, where $\boldsymbol{\sigma}(\boldsymbol{p}) = \prod_{i \in [n]} \sigma_i (p_i)$ and $p_i$ is the corresponding strategy of player $i$ in $\boldsymbol{p}$.
The (pure) optimal load assignment denoted by $\boldsymbol{o}$ minimizes social delay among all possible (pure) load assignments which might be in contrast with the selfish behavior of players.
The (pure) \textit{price of anarchy} (PoA) of a congestion game is the maximum ratio $D(\boldsymbol{p}) \slash D(\boldsymbol{o})$ over all equilibria $\boldsymbol{p}$ of the game.
Throughout the article, we follow the convention that $y \leq \boldsymbol{x}$ means that $y$ is less than or equal to all elements of the vector $\boldsymbol{x}$.

\section{Risk-Averse Equilibrium for Stochastic Congestion Games}
\label{section_risk_averse_equilibrium_congestion_game}

In the following sub-section, illustrative examples are provided with analysis of their equilibria in classic and risk-averse frameworks which motivate the novel risk-averse approach for stochastic congestion games presented in this article.

\subsection{Illustrative Examples}
\label{sub_section_illustrative_example}
The Pigou network \citep{pigou2013economics} is one of the simplest networks studied in congestion games. We first use the Pigou network to clearly state the motivation of the current work in the first example. We then study the more controversial network used by 
%Braess 
\cite{braess1968paradoxon} in the famous Braess's paradox in the second example. The two examples below set grounding for the risk-averse equilibrium for congestion games proposed in this article.

\begin{example}
\label{example1}
Consider the Pigou network with two parallel links between source and destination as shown in Figure \ref{figure_example1_Pigou}.
There are $n$ players (vehicles or data packages) to travel from source to destination.
The top and bottom links are labeled as $1$ and $2$ with loads $m_1$ and $m_2 = n - m_1$, respectively. The travel times on links $1$ and $2$ are respectively independent random variables $L_1(m_1)$ and $L_2(m_2)$ with expected values $l_1(m_1) = \frac{m_1}{n}$ and $l_2(m_2) = 1$ and pdfs
\[
\begin{aligned}
    f_1(x | m_1) = \ & \alpha \Bigg ( 2exp \bigg ( -100 \Big (x - \frac{m_1}{4n} \Big )^2 \bigg ) \cdot \mathbbm{1} \left \{ 0 \leq x \leq \frac{m_1}{2n} \right \} \\
    & \ \ \ \ + 3exp \bigg ( -100 \Big (x - \frac{3m_1}{2n} \Big )^2 \bigg ) \cdot \mathbbm{1} \left \{ \frac{5m_1}{4n} \leq x \leq \frac{7m_1}{4n} \right \}  \Bigg ), \\
    f_2(x | m_2) = \ & \beta exp \left ( -100 \left (x - 1 \right )^2 \right ) \cdot \mathbbm{1} \left \{ \frac34 \leq x \leq \frac54 \right \},
\end{aligned}
\]
where $\alpha$ and $\beta$ are constants for which each of the two distributions integrate to one and $\mathbbm{1}\{.\}$ is the indicator function.

\begin{figure}[t]
\centering
\includegraphics[width=0.5\textwidth]{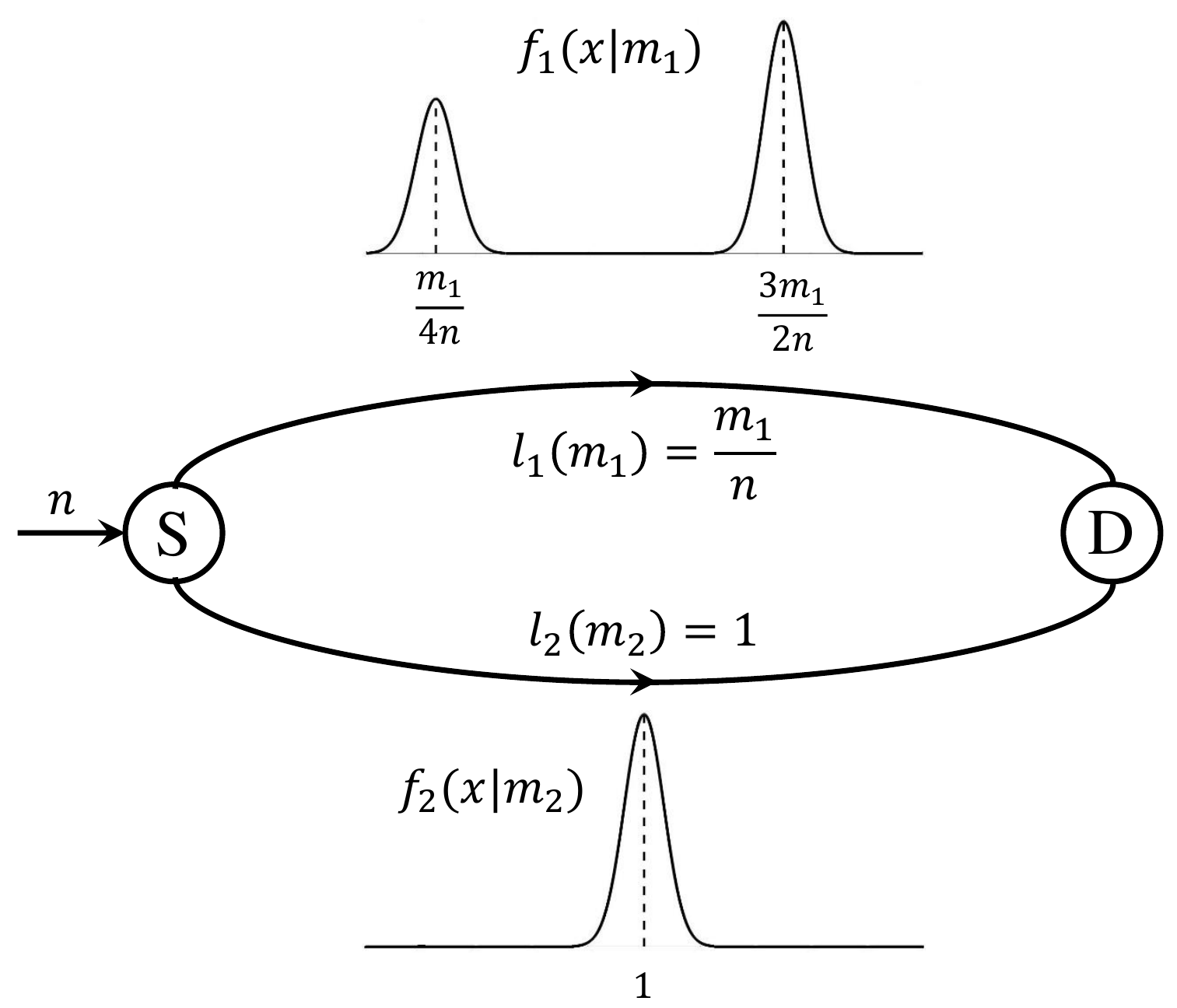}
\caption{The Pigou network in Example \ref{example1} with the load-dependent latency pdfs and the corresponding means of links.}
\label{figure_example1_Pigou}
\end{figure}

\end{example}

The well-known Wardrop equilibrium \citep{wardrop1952road, wardrop1952correspondence}, also Nash equilibrium \citep{vonNeumann1947}, for the Pigou network in Example \ref{example1} is that all the $n$ players travel along the top link since it is the weakly dominant strategy for any player as the expected latency incurred along the top link is always less than or equal to the expected latency incurred along the bottom link, $l_1(m_1) = \frac{m_1}{n} \leq 1 = l_2(m_2)$.
As a result, the Wardrop equilibrium for Pigou network is $\boldsymbol{p}_W^* = (1, 1, \dots, 1)$ with social delay $D_W(\boldsymbol{p}_W^*) = 1$.
However, although the expected latency along the top link is less than or equal to that of the bottom link, $l_1(m_1) \leq l_2(m_2)$, the variance of travel time along the top link at full capacity is larger than that along the bottom link, which increases the risk and uncertainty of traveling along the top link \citep{yekkehkhany2020risk, yekkehkhanycost, 9142286}.
In fact, the bottom link with higher expected travel time is more likely to have a lower delay than the top link at full capacity, i.e., $P \big ( L_2(0) \leq L_1(n)  \big ) = 0.6 > 0.5$.
As a result, a risk-averse player selects the bottom link for commute when the top link is at full capacity, especially if it is a one-time trip. We will also shown later, the risk-averse behavior of players decreases social delay for this example.
As an example, consider a traveler who wants to go from hotel to airport who has two options for this trip: taking the highway that has lower expected travel time, but is more likely to get congested due to traffic jams and crashes (top link in Pigou network), or taking the urban streets with a higher expected travel time and lower congestion (the bottom link in Pigou network).
A risk-neutral player travels along the top link with lower expected latency, but a risk-averse player travels along the bottom link to assure not to incur a long delay and miss the flight.
Even in everyday commutes between home and work, the expected delay over many days may not be a desirable objective to minimize. No-one desires to arrive early to work some days but late on others, and to be penalized accordingly.
The Braess network, studied in the next example, further enforces the fact that minimizing the expected delay is not desirable for risk-averse players.

\begin{figure}[t]
\centering
\includegraphics[width=0.75\textwidth]{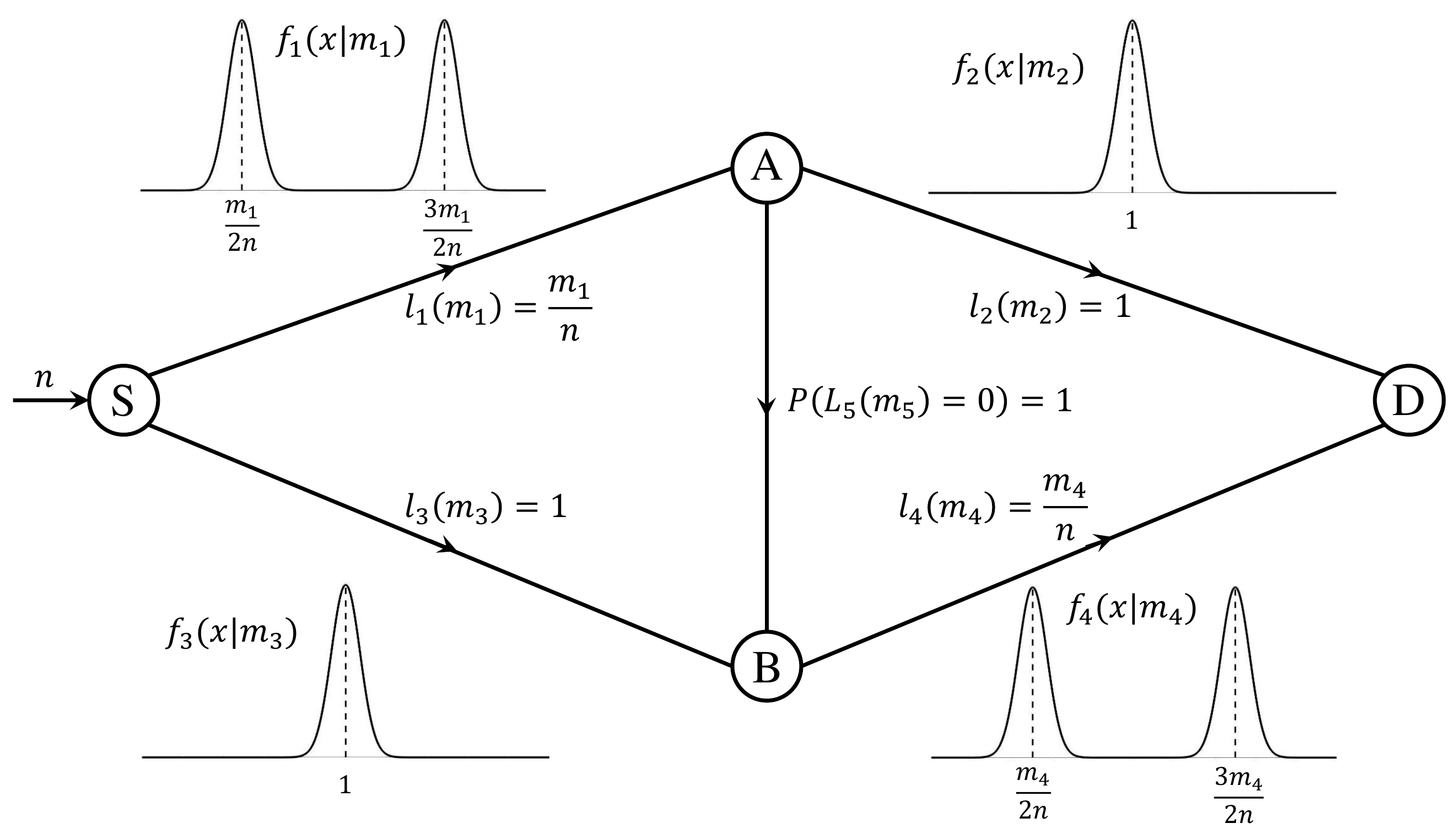}
\caption{The Braess network in Example \ref{example2} with the load-dependent latency pdfs and the corresponding means of links.}
\label{figure_example2_Braess}
\end{figure}

\begin{example}
\label{example2}
Consider the Braess network depicted in Figure \ref{figure_example2_Braess}.
There are $n$ players (vehicles or data packages) to travel from source to destination.
Other than the source and destination, there are two nodes $A$ and $B$ in the network.
The directed links $(S, A)$, $(A, D)$, $(S, B)$, $(B, D)$, and $(A, B)$ are referred to as links $1$, $2$, $3$, $4$, and $5$ with loads $m_1$, $m_2$, $m_3$, $m_4$, and $m_5$, respectively.
The travel times on links $1$, $2$, $3$, $4$, and $5$ are respectively independent random variables $L_1(m_1)$, $L_2(m_2)$, $L_3(m_3)$, $L_4(m_4)$, and $L_5(m_5)$ with expected values $l_1(m_1) = \frac{m_1}{n}$, $l_2(m_2) = 1$, $l_3(m_3) = 1$, $l_4(m_4) = \frac{m_4}{n}$, and $l_5(m_5) = 0$ and pdfs
\[
\begin{aligned}
    f_1(x | m_1) = \ & \gamma \Bigg ( exp \bigg ( -100 \Big (x - \frac{m_1}{2n} \Big )^2 \bigg ) \cdot \mathbbm{1} \left \{ 0 \leq x \leq \frac{m_1}{n} \right \} \\
    & \ \ \ \ + exp \bigg ( -100 \Big (x - \frac{3m_1}{2n} \Big )^2 \bigg ) \cdot \mathbbm{1} \left \{ \frac{m_1}{n} < x \leq \frac{2m_1}{n} \right \}  \Bigg ), \\
    f_2(x | m_2) = \ & \zeta exp \left ( -100 \left (x - 1 \right )^2 \right ) \cdot \mathbbm{1} \left \{ \frac12 \leq x \leq \frac32 \right \}, \\
    f_3(x | m_3) = \ & \zeta exp \left ( -100 \left (x - 1 \right )^2 \right ) \cdot \mathbbm{1} \left \{ \frac12 \leq x \leq \frac32 \right \}, \\
    f_4(x | m_4) = \ & \gamma \Bigg ( exp \bigg ( -100 \Big (x - \frac{m_4}{2n} \Big )^2 \bigg ) \cdot \mathbbm{1} \left \{ 0 \leq x \leq \frac{m_4}{n} \right \} \\
    & \ \ \ \ + exp \bigg ( -100 \Big (x - \frac{3m_4}{2n} \Big )^2 \bigg ) \cdot \mathbbm{1} \left \{ \frac{m_4}{n} < x \leq \frac{2m_4}{n} \right \}  \Bigg ),
\end{aligned}
\]
where $\gamma$ and $\zeta$ are constants for which the distributions integrate to one, $\mathbbm{1}\{.\}$ is the indicator function, and $P \big ( L_5(m_5) = 0 \big ) = 1$.
There are three paths from source to destination, $(S, A, D)$, $(S, A, B, D)$, and $(S, B, D)$, that are referred to as paths $1, 2$, and $3$ with loads $m^1, m^2$, and $m^3$, respectively, where the difference between links and paths should be clear from the context.
Note that the link loads are related to path loads as $m_1 = m^1 + m^2$, $m_2 = m^1$, $m_3 = m^3$, $m_4 = m^2 + m^3$, and $m_5 = m^2$, and the delays along paths are related to link delays as $L^1(\boldsymbol{m}) = L_1(m_1) + L_2(m_2)$, $L^2(\boldsymbol{m}) = L_1(m_1) + L_5(m_5) + L_4(m_4) = L_1(m_1) + L_4(m_4)$, and $L^3(\boldsymbol{m}) = L_3(m_3) + L_4(m_4)$.

\end{example}

The Wardrop (Nash) equilibrium for the Braess network in Example \ref{example2} is that all the $n$ players travel along path $2$ since it is the weakly dominant path for any player as the expected latency incurred along path $2$ is always less than or equal to the expected latency incurred along the other two paths $1$ and $3$,
\[
l^2(\boldsymbol{m}) = l_1(m_1) + l_5(m_5) + l_4(m_4) = \frac{m_1}{n} + \frac{m_4}{n}
\begin{cases}
\leq \frac{m_1}{n} + 1 = l_1(m_1) + l_2(m_2) = l^1(\boldsymbol{m}), \\
\leq 1 + \frac{m_4}{n} = l_3(m_3) + l_4(m_4) = l^3(\boldsymbol{m}).
\end{cases}
\]
As a result, the Wardrop equilibrium for Braess network is $\boldsymbol{p}_W^* = (2, 2, \dots, 2)$ with social delay $D_W(\boldsymbol{p}_W^*) = 2$.
However, although path $2$ has latency less than or equal to that of paths $1$ and $3$, $l^2(\boldsymbol{m}) \leq \big (l^1(\boldsymbol{m}), l^3(\boldsymbol{m}) \big)$, the variance of travel time along path $2$ at full capacity is larger than that along paths $1$ and $3$, which increases the risk and uncertainty of traveling along path $2$.
In fact, path $1$ (or $3$) with higher expected travel time is more likely to have a lower delay than the rest of the paths, i.e., $P \Big ( L^1(0) \leq \big ( L^2(n), L^3(0) \big )  \Big ) = \frac38 > \frac14 = P \Big ( L^2(n) \leq \big ( L^1(0), L^3(0) \big )  \Big )$.
As a result, a risk-averse player selects paths $1$ or $3$ for commute when path $2$ is at full capacity, and as is shown later, the risk-averse behavior of players decreases social delay for this example.

\subsection{Risk-Averse Equilibrium}
\label{sub_section_RAE}
In the classical Wardrop (Nash) equilibrium, the best response of player $i \in [n]$ to the mixed strategy $\boldsymbol{\sigma}_{-i}$ of the other $[n] \setminus i$ players is defined as the set
\[
\underset{p_i \in \mathcal{P}_i}{\argmin} \ \overline{l}^{i}(p_i, \boldsymbol{\sigma}_{-i}).
\]
In other words, the best response for player $i$ given $\boldsymbol{\sigma}_{-i}$ is defined as the path that minimizes the expected travel time.
However, motivated by Examples \ref{example1} and \ref{example2}, the path with minimum expected latency may have a high volatility as well that causes risky scenarios for travelers.
As a result, the classical Wardrop (Nash) equilibrium that ignores the distribution of path latency except for taking the expected latency into account, that does not carry any information about variance and the shape of the distribution, falls short in addressing risk-averse behavior of players.
In this article, motivated by Examples \ref{example1} and \ref{example2}, we propose a Risk-Averse Best Action Decision with Incomplete Information (R-ABADI) of a player to the strategy of the other players in a stochastic congestion game as follows.

\begin{definition}
\label{def_best_response_mixed_RAE}
Given the mixed strategy profile $\boldsymbol{\sigma}_{-i}$ of players $[n] \setminus i$, the set of mixed strategy risk-averse best responses of player $i$ is the set of all probability distributions over the set
\begin{equation}
    \label{eq_mixed_best_response}
    \underset{p_i \in \mathcal{P}_i}{\argmax} \ P \left ( \overline{L}^{i}(p_i, \boldsymbol{\sigma}_{-i}) \leq \overline{\boldsymbol{L}}^{i}(\mathcal{P}_i \setminus p_i, \boldsymbol{\sigma}_{-i}) \right ),
\end{equation}
where what we mean by $\overline{L}^{i}(p_i, \boldsymbol{\sigma}_{-i})$ being less than or equal to $\overline{\boldsymbol{L}}^{i}(\mathcal{P}_i \setminus p_i, \boldsymbol{\sigma}_{-i})$ when $\mathcal{P}_i \setminus p_i \neq \oldemptyset$ is that  $\overline{L}^{i}(p_i, \boldsymbol{\sigma}_{-i})$ is less than or equal to $\overline{L}^{i}(p_i', \boldsymbol{\sigma}_{-i})$ for all $p_i' \in \mathcal{P}_i \setminus p_i$; otherwise, if $\mathcal{P}_i \setminus p_i = \oldemptyset$, player $i$ only has a single option that can be played.
The same randomness on the action of players $[n] \setminus i$ is considered in $\overline{L}^{i}(p_i, \boldsymbol{\sigma}_{-i})$ for all $p_i \in \mathcal{P}_i$.
Given the mixed strategy $\boldsymbol{\sigma}_{-i}$ of players $[n] \setminus i$, the risk-averse best response set of player $i$'s strategies is denoted by $RB(\boldsymbol{\sigma}_{-i})$, which is in general a set-valued function.
\end{definition}

The risk-averse equilibrium for stochastic congestion games is defined as follows.

\begin{definition}
\label{def_mixed_strategy_RAE}
A strategy profile $\boldsymbol{\sigma}^* = (\sigma_1^*, \sigma_2^*, \dots,$ $\sigma_N^*)$ is a risk-averse equilibrium if and only if $\sigma_i^* \in RB(\boldsymbol{\sigma}_{-i}^*)$ for all $i \in [n]$.
\end{definition}

The following theorem proves the existence of a risk-averse equilibrium for any stochastic congestion game with finite number of players and pure strategy sets $\mathcal{P}_i$ for all $i \in [n]$ with finite cardinality.

\begin{theorem}
\label{theorem_existence_RAE}
For any finite $n$-player stochastic congestion game, a risk-averse equilibrium exists.
\end{theorem}

The proof of Theorem \ref{theorem_existence_RAE} is provided in Appendix \ref{proof_theorem_RAE}.

As a direct result of Definitions \ref{def_best_response_mixed_RAE} and \ref{def_mixed_strategy_RAE}, the pure strategy risk-averse best response and pure strategy risk-averse equilibrium are defined as follows. The pure strategy risk-averse best response of player $i$ to the pure strategy $\boldsymbol{p}_{-i}$ of players $[n] \setminus i$ is the set
\begin{equation}
    \label{pure_best_response}
    \left\{
    \begin{array}{ll}
        \argmax_{p_i \in \mathcal{P}_i} P \Big ( L^{i} \left ( p_i, \boldsymbol{p}_{-i} \right ) \leq \boldsymbol{L}^{i} \left ( \mathcal{P}_i \setminus p_i, \boldsymbol{p}_{-i} \right ) \Big ), \ \ \ \text{ if } \mathcal{P}_i \setminus p_i \neq \oldemptyset,
        \\
        p_i, \ \ \  \text{ if } \mathcal{P}_i \setminus p_i = \oldemptyset.
    \end{array}
    \right.
\end{equation}
Given the pure strategy $\boldsymbol{p}_{-i}$ of players $[n] \setminus i$, the risk-averse best response set of player $i$ in Equation \eqref{pure_best_response} is denoted by $RB(\boldsymbol{p}_{-i})$ (overloading notation, $RB(.)$ is used for both pure and mixed strategy risk-averse best responses).
As a result, a pure strategy profile $\boldsymbol{p}^* = (p_1^*, p_2^*, \dots, p_n^*)$ is a pure strategy risk-averse equilibrium if and only if $p_i^* \in RB(\boldsymbol{p}_{-i}^*)$ for all $i \in [n]$.

Strict dominance in the classical Wardrop (Nash) equilibrium is defined as follows.
A pure strategy $p_i \in \mathcal{P}_i$ of player $i$ strictly dominates a second pure strategy $p_i' \in \mathcal{P}_i$ of the player if
\[
l^{i}(p_i, \boldsymbol{p}_{-i}) < l^{i}(p_i', \boldsymbol{p}_{-i}), \ \forall \boldsymbol{p}_{-i} \in \boldsymbol{\mathcal{P}}_{-i}.
\]
The solution concept of iterated elimination of strictly dominated strategies can also be applied to the risk-averse equilibrium using the following definition.

\begin{definition}
\label{def_strict_dominance_RAE}
A pure strategy $p_i \in \mathcal{P}_i$ of player $i$ strictly dominates a second pure strategy $p_i' \in \mathcal{P}_i$ of the player in the risk-averse equilibrium if
\begin{equation}
    \label{eq_strict_dominance_RAE}
    P \Big ( L^{i} \left ( p_i, \boldsymbol{p}_{-i} \right ) \leq \boldsymbol{L}^{i} \left ( \mathcal{P}_i \setminus p_i, \boldsymbol{p}_{-i} \right ) \Big ) > P \Big ( L^{i} \left ( p_i', \boldsymbol{p}_{-i} \right ) \leq \boldsymbol{L}^{i} \left ( \mathcal{P}_i \setminus p_i', \boldsymbol{p}_{-i} \right ) \Big ), \ \forall \boldsymbol{p}_{-i} \in \boldsymbol{\mathcal{P}}_{-i}.
\end{equation}
\end{definition}

Consider path $p_i \in \mathcal{P}_i$ strictly dominates path $p_i' \in \mathcal{P}_i$ for player $i$; then, for any $\boldsymbol{\sigma}_{-i} \in \boldsymbol{\Sigma}_{-i}$
\begin{equation}
    \label{eq_proof_strict_dominance_RAE}
    \begin{aligned}
        & P \left ( \overline{L}^{i}(p_i, \boldsymbol{\sigma}_{-i}) \leq \overline{\boldsymbol{L}}^{i}(\mathcal{P}_i \setminus p_i, \boldsymbol{\sigma}_{-i}) \right ) \\
        \overset{(a)}{=} & \sum_{\boldsymbol{p}_{-i} \in \boldsymbol{\mathcal{P}}_{-i}} \Bigg ( P \left ( L^{i}(p_i, \boldsymbol{p}_{-i}) \leq \boldsymbol{L}^{i}(\mathcal{P}_i \setminus p_i, \boldsymbol{p}_{-i}) \right ) \cdot \boldsymbol{\sigma}(\boldsymbol{p}_{-i}) \Bigg ) \\
        \overset{(b)}{>} & \sum_{\boldsymbol{p}_{-i} \in \boldsymbol{\mathcal{P}}_{-i}} \Bigg ( P \left ( L^{i}(p_i', \boldsymbol{p}_{-i}) \leq \boldsymbol{L}^{i}(\mathcal{P}_i \setminus p_i', \boldsymbol{p}_{-i}) \right ) \cdot \boldsymbol{\sigma}(\boldsymbol{p}_{-i}) \Bigg ) \\
        = & \ P \left ( \overline{L}^{i}(p_i', \boldsymbol{\sigma}_{-i}) \leq \overline{\boldsymbol{L}}^{i}(\mathcal{P}_i \setminus p_i', \boldsymbol{\sigma}_{-i}) \right ),
    \end{aligned}
\end{equation}
where $(a)$ is true by the law of total probability, $\boldsymbol{\sigma}(\boldsymbol{p}_{-i}) =  \prod_{j \in [n] \setminus i} \sigma_j (p_j)$ and $p_j$ is the corresponding strategy of player $j$ in $\boldsymbol{p}_{-i}$, and
$(b)$ is followed by Equation \eqref{eq_strict_dominance_RAE} in Definition \ref{def_strict_dominance_RAE}.
By Equation \eqref{eq_proof_strict_dominance_RAE} and Equation \eqref{eq_mixed_best_response} in Definition \ref{def_best_response_mixed_RAE}, a strictly dominated pure strategy cannot be a best response to any mixed strategy profile $\boldsymbol{\sigma}_{-i} \in \boldsymbol{\Sigma}_{-i}$, so it can be removed from the set of strategies of player $i$.

In order to find the risk-averse equilibrium for a stochastic congestion game, we use support enumeration.
For example, hypothesize that $\boldsymbol{\mathcal{P}}' \coloneqq \{ \mathcal{P}_1', \mathcal{P}_2', \dots, \mathcal{P}_n' \}$ is the \textit{support} of a risk-averse equilibrium, where $\mathcal{P}_i'$ is the set of pure strategies of player $i$ that are played with non-zero probability and $\sigma_i(p_i)$ for $p_i \in \mathcal{P}_i'$ indicates the probability mass function on the support.
At equilibrium, player $i \in [n]$ should be indifferent between strategies in the set $\mathcal{P}_i'$, has no incentive to deviate to the rest of strategies in the set $\mathcal{P}_i \setminus \mathcal{P}_i'$, and the probability mass function over the support should add to one.
As a result, if there is a risk-averse equilibrium with the mentioned support, it is the solution of the following set of equations for $\boldsymbol{\sigma} \in \boldsymbol{\Sigma}$:
\begin{equation}
\label{eq_find_mixed_strategy_RAE}
    \left\{
    \begin{array}{ll}
        P \left ( \overline{L}^{i}(p_i', \boldsymbol{\sigma}_{-i}) \leq \overline{\boldsymbol{L}}^{i}(\mathcal{P}_i \setminus p_i', \boldsymbol{\sigma}_{-i}) \right ) \geq P \left ( \overline{L}^{i}(p_i, \boldsymbol{\sigma}_{-i}) \leq \overline{\boldsymbol{L}}^{i}(\mathcal{P}_i \setminus p_i, \boldsymbol{\sigma}_{-i}) \right ), \forall p_i\in \mathcal{P}_i, p_i' \in \mathcal{P}_i', \forall i \in [n],\\
        \\
        \sum_{p_i \in \mathcal{P}_i'} \sigma_i(p_i) = 1, \forall i \in [n],\\
        \\
        \sigma_i(p_i) = 0, \forall p_i \in \mathcal{P}_i \setminus \mathcal{P}_i', \forall i \in [n].
    \end{array}
    \right.
\end{equation}

As mentioned earlier in Equation \eqref{eq_proof_strict_dominance_RAE}, using the law of total probability, we have
\begin{equation}
\label{eq_LTP_RAE}
\begin{aligned}
    & P \left ( \overline{L}^{i}(p_i, \boldsymbol{\sigma}_{-i}) \leq \overline{\boldsymbol{L}}^{i}(\mathcal{P}_i \setminus p_i, \boldsymbol{\sigma}_{-i}) \right ) \\
    = & \sum_{\boldsymbol{p}_{-i} \in \boldsymbol{\mathcal{P}}_{-i}} \Bigg ( P \left ( L^{i}(p_i, \boldsymbol{p}_{-i}) \leq \boldsymbol{L}^{i}(\mathcal{P}_i \setminus p_i, \boldsymbol{p}_{-i}) \right ) \cdot \boldsymbol{\sigma}(\boldsymbol{p}_{-i}) \Bigg ) \\
    = & \sum_{\boldsymbol{p}_{-i} \in \boldsymbol{\mathcal{P}}_{-i}} \Big ( t_i(p_i, \boldsymbol{p}_{-i}) \cdot \boldsymbol{\sigma}(\boldsymbol{p}_{-i}) \Big ),
\end{aligned}
\end{equation}
where $t_i(p_i, \boldsymbol{p}_{-i}) \coloneqq P \left ( L^{i}(p_i, \boldsymbol{p}_{-i}) \leq \boldsymbol{L}^{i}(\mathcal{P}_i \setminus p_i, \boldsymbol{p}_{-i}) \right )$ is the $i$-th element of an $n$-dimensional vector called $\boldsymbol{t}(p_i, \boldsymbol{p}_{-i})$.
Construct a \textit{risk-averse probability tensor} of rank $n$ where $\mathcal{P}_i$ forms the $i$-th dimension of the tensor.
Let the element associated with $(p_i, \boldsymbol{p}_{-i})$ in the tensor be the vector $\boldsymbol{t}(p_i, \boldsymbol{p}_{-i})$.
Equations \eqref{eq_find_mixed_strategy_RAE} and \eqref{eq_LTP_RAE} along with the definition of the risk-averse probability tensor provide us with an alternative approach for deriving the risk-averse equilibrium, which is to find the Wardrop (Nash) equilibrium on the risk-averse probability tensor.

The mean-variance (MV) and conditional value at risk level $\alpha$ (CVaR$_\alpha$) methods are two well-known frameworks to consider risk in statistics.
In the next two sub-sections, two new risk-averse equilibria based on these two concepts are proposed.

\subsection{Mean-Variance Equilibrium}
\label{sec_mean_variance_equilibrium}
As seen in Examples \ref{example1} and \ref{example2}, the high variance of paths with lower expected travel time can result in uncertainty and impose high latency for travelers.
The mean-variance framework in statistics addresses this issue by keeping a balance between low latency and low variance.
Applying this method to the proposed stochastic congestion game setting, the mean-variance best response and mean-variance equilibrium are defined as follows.

\begin{definition}
\label{def_best_response_mixed_MV}
Given the mixed strategy profile $\boldsymbol{\sigma}_{-i}$ of players $[n] \setminus i$, the set of mixed strategy mean-variance best responses of player $i$ is the set of all probability distributions over the set
\begin{equation}
    \label{eq_mixed_best_response_MV}
    \underset{p_i \in \mathcal{P}_i}{\argmin} \ \mathrm{Var} \left (\overline{L}^{i}(p_i, \boldsymbol{\sigma}_{-i}) \right ) + \rho \cdot \overline{l}^{i}(p_i, \boldsymbol{\sigma}_{-i}),
\end{equation}
where the variance $\mathrm{Var} \left (\overline{L}^{i}(p_i, \boldsymbol{\sigma}_{-i}) \right )$ can be calculated using the pdf of $\overline{L}^{i}(p_i, \boldsymbol{\sigma}_{-i})$ provided in Equation \eqref{dist_mixed} and $\rho \geq 0$ is a hyper-parameter capturing the absolute risk tolerance.
Given the mixed strategy $\boldsymbol{\sigma}_{-i}$ of players $[n] \setminus i$, the mean-variance best response set of player $i$'s strategies is denoted by $MB(\boldsymbol{\sigma}_{-i})$, which is in general a set-valued function.

\end{definition}

\begin{definition}
\label{def_mixed_strategy_MV}
A strategy profile $\boldsymbol{\sigma}^* = (\sigma_1^*, \sigma_2^*, \dots,$ $\sigma_N^*)$ is a mean-variance equilibrium if and only if $\sigma_i^* \in MB(\boldsymbol{\sigma}_{-i}^*)$ for all $i \in [n]$.
\end{definition}

The existence of the mean-variance equilibrium is discussed in the following theorem.

\begin{theorem}
\label{theorem_existence_MV}
For any finite $n$-player stochastic congestion game, a mean-variance equilibrium exists.
\end{theorem}

The proof of Theorem \ref{theorem_existence_MV} is provided in Appendix \ref{proof_theorem_MV}.

The pure strategy mean-variance best response of player $i$ to the pure strategy $\boldsymbol{p}_{-i}$ of players $[n] \setminus i$ is the set
\begin{equation}
    \label{pure_best_response_MV}
    \underset{p_i \in \mathcal{P}_i}{\argmin} \ \mathrm{Var} \left (L^{i}(p_i, \boldsymbol{p}_{-i}) \right ) + \rho \cdot l^{i}(p_i, \boldsymbol{p}_{-i}),
\end{equation}
where $\mathrm{Var} \left (L^{i}(p_i, \boldsymbol{p}_{-i}) \right ) = \mathrm{Var} \left ( \sum_{e \in p_i} L_e(p_i, \boldsymbol{p}_{-i}) \right ) = \sum_{e \in p_i} \sum_{e' \in p_i} \mathrm{Cov} \left ( L_e(p_i, \boldsymbol{p}_{-i}), L_{e'}(p_i, \boldsymbol{p}_{-i}) \right )$.
Given the pure strategy $\boldsymbol{p}_{-i}$ of players $[n] \setminus i$, the mean-variance best response set of player $i$ in Equation \eqref{pure_best_response_MV} is denoted by $MB(\boldsymbol{p}_{-i})$ (overloading notation, $MB(.)$ is used for both pure and mixed strategy mean-variance best responses).
As a result, a pure strategy profile $\boldsymbol{p}^* = (p_1^*, p_2^*, \dots, p_n^*)$ is a pure strategy mean-variance equilibrium if and only if $p_i^* \in MB(\boldsymbol{p}_{-i}^*)$ for all $i \in [n]$.
The strict dominance concept is straightforward among pure strategy profiles in mean-variance equilibrium that is defined as follows.
A pure strategy $p_i \in \mathcal{P}_i$ of player $i$ strictly dominates a second pure strategy $p_i' \in \mathcal{P}_i$ of the player in pure strategy mean-variance equilibrium if
\begin{equation}
    \label{eq_strict_dominance_MV}
    \mathrm{Var} \left (L^{i}(p_i, \boldsymbol{p}_{-i}) \right ) + \rho \cdot l^{i}(p_i, \boldsymbol{p}_{-i}) < \mathrm{Var} \left (L^{i}(p_i', \boldsymbol{p}_{-i}) \right ) + \rho \cdot l^{i}(p_i', \boldsymbol{p}_{-i}), \ \forall \boldsymbol{p}_{-i} \in \boldsymbol{\mathcal{P}}_{-i}.
\end{equation}
%However, since $\mathrm{Var} \left (\overline{L}^{p_i}(p_i, \boldsymbol{\sigma}_{-i}) \right )$ has a complicated relation with $\mathrm{Var} \left (L^{p_i}(p_i, \boldsymbol{p}_{-i}) \right )$ for all $\boldsymbol{p}_{-i} \in \boldsymbol{\mathcal{P}}_{-i}$ and $\boldsymbol{\sigma}(\boldsymbol{p}_{-i})$, and the relation is not in a simple linear form in terms of $\boldsymbol{\sigma}(\boldsymbol{p}_{-i})$ as variance is not a linear operator, the strict dominance concept in general is complicated and does not necessarily simplify the process of finding the mean-variance equilibrium.
However, due to the fact that variance is not a linear operator, strict dominance may not be derived from Equation \eqref{eq_strict_dominance_MV} for mixed strategy mean-variance equilibrium as described below.
\begin{equation}
    \label{eq_strict_dominance_MV_general}
    \begin{aligned}
    & \mathrm{Var} \left (\overline{L}^{i}(p_i, \boldsymbol{\sigma}_{-i}) \right ) + \rho \cdot \overline{l}^{i}(p_i, \boldsymbol{\sigma}_{-i}) \\
    \overset{(a)}{=} & \ \mathrm{E} \left [ \left ( \overline{L}^{i}(p_i, \boldsymbol{\sigma}_{-i}) \right)^2 \right ] - \left ( \overline{l}^{i}(p_i, \boldsymbol{\sigma}_{-i}) \right )^2 + \rho \cdot \overline{l}^{i}(p_i, \boldsymbol{\sigma}_{-i}) \\
    \overset{(b)}{=} & \sum_{\boldsymbol{p}_{-i} \in \boldsymbol{\mathcal{P}}_{-i}} \left ( \boldsymbol{\sigma}(\boldsymbol{p}_{-i}) \cdot \mathrm{E} \left [ \left ( L^{i}(p_i, \boldsymbol{p}_{-i}) \right)^2 \right ] \right ) - \left ( \sum_{\boldsymbol{p}_{-i} \in \boldsymbol{\mathcal{P}}_{-i}} \Big ( \boldsymbol{\sigma}(\boldsymbol{p}_{-i}) \cdot l^{i}(p_i, \boldsymbol{p}_{-i}) \Big ) \right )^2 \\
    & + \rho \sum_{\boldsymbol{p}_{-i} \in \boldsymbol{\mathcal{P}}_{-i}} \Big ( \boldsymbol{\sigma}(\boldsymbol{p}_{-i}) \cdot l^{i}(p_i, \boldsymbol{p}_{-i}) \Big ) \\
    \overset{(c)}{=} & \sum_{\boldsymbol{p}_{-i} \in \boldsymbol{\mathcal{P}}_{-i}} \left ( \boldsymbol{\sigma}(\boldsymbol{p}_{-i}) \cdot \mathrm{E} \left [ \left ( L^{i}(p_i, \boldsymbol{p}_{-i}) \right)^2 \right ] \right ) \\
    & - \sum_{\boldsymbol{p}_{-i} \in \boldsymbol{\mathcal{P}}_{-i}} \left ( \sum_{\boldsymbol{p}_{-i}' \in \boldsymbol{\mathcal{P}}_{-i}} \Big ( \boldsymbol{\sigma}(\boldsymbol{p}_{-i}) \cdot \boldsymbol{\sigma}(\boldsymbol{p}_{-i}') \cdot l^{i}(p_i, \boldsymbol{p}_{-i}) \cdot l^{i}(p_i, \boldsymbol{p}_{-i}') \Big ) \right ) + \rho \cdot \sum_{\boldsymbol{p}_{-i} \in \boldsymbol{\mathcal{P}}_{-i}} \Big ( \boldsymbol{\sigma}(\boldsymbol{p}_{-i}) \cdot l^{i}(p_i, \boldsymbol{p}_{-i}) \Big ) \\
    \overset{(d)}{=} & \sum_{\boldsymbol{p}_{-i} \in \boldsymbol{\mathcal{P}}_{-i}} \boldsymbol{\sigma}(\boldsymbol{p}_{-i}) \cdot \left ( \mathrm{E} \left [ \left ( L^{i}(p_i, \boldsymbol{p}_{-i}) \right)^2 \right ] - l^{i}(p_i, \boldsymbol{p}_{-i}) \cdot \sum_{\boldsymbol{p}_{-i}' \in \boldsymbol{\mathcal{P}}_{-i}} \Big ( \boldsymbol{\sigma}(\boldsymbol{p}_{-i}') \cdot l^{i}(p_i, \boldsymbol{p}_{-i}') \Big ) + \rho \cdot l^{i}(p_i, \boldsymbol{p}_{-i}) \right ) \\
    = & \sum_{\boldsymbol{p}_{-i} \in \boldsymbol{\mathcal{P}}_{-i}} \boldsymbol{\sigma}(\boldsymbol{p}_{-i}) \cdot \left ( \mathrm{E} \left [ \left ( L^{i}(p_i, \boldsymbol{p}_{-i}) \right)^2 \right ] - l^{i}(p_i, \boldsymbol{p}_{-i}) \cdot \left ( \sum_{\boldsymbol{p}_{-i}' \in \boldsymbol{\mathcal{P}}_{-i}} \Big ( \boldsymbol{\sigma}(\boldsymbol{p}_{-i}') \cdot l^{i}(p_i, \boldsymbol{p}_{-i}') \Big ) + \rho \right ) \right ),
    \end{aligned}
\end{equation}
where $(a)$ is true by the definition of variance, $(b)$ is followed by Equation \eqref{eq_mean_mixed}, $(c)$ is derived by expanding the second term, and $(d)$ is true by combining the summation over $\boldsymbol{p}_{-i} \in \boldsymbol{\mathcal{P}}_{-i}$ and factoring $\boldsymbol{\sigma}(\boldsymbol{p}_{-i})$.
As can be seen in Equation \eqref{eq_strict_dominance_MV_general}, since variance is a non-linear operator, it is not clear whether Equation \eqref{eq_strict_dominance_MV} can result in $\mathrm{Var} \left (\overline{L}^{i}(p_i, \boldsymbol{\sigma}_{-i}) \right ) + \rho \cdot \overline{l}^{i}(p_i, \boldsymbol{\sigma}_{-i}) < \mathrm{Var} \left (\overline{L}^{i}(p_i', \boldsymbol{\sigma}_{-i}) \right ) + \rho \cdot \overline{l}^{i}(p_i', \boldsymbol{\sigma}_{-i})$ for all $\boldsymbol{\sigma}_{-i} \in \boldsymbol{\Sigma}_{-i}$.
As a result, use of strict dominance in the mixed strategy mean-variance equilibrium is not advised.
%The complication of the strict dominance is 
In certain circumstances though, we can propose conditions for strict dominance; e.g., when $l^{i}(\boldsymbol{p}) \leq \frac{\rho}{2}$ for all $\boldsymbol{p} \in \boldsymbol{\mathcal{P}}$ and for all $i \in [n]$ which is discussed in the following definition or when $l^{i}(\boldsymbol{p}) \geq \frac{\rho}{2}$ for all $\boldsymbol{p} \in \boldsymbol{\mathcal{P}}$ and for all $i \in [n]$.

\begin{definition}
\label{def_strict_dominance_MV_new}
If $l^{i}(\boldsymbol{p}) \leq \frac{\rho}{2}$ for all $\boldsymbol{p} \in \boldsymbol{\mathcal{P}}$ and for all $i \in [n]$, pure strategy $p_i \in \mathcal{P}_i$ of player $i$ strictly dominates a second pure strategy $p_i' \in \mathcal{P}_i$ of the player in the mean-variance equilibrium if
\begin{equation}
    \label{eq_strict_dominance_MV_sufficient_mean}
    l^{i} \left ( p_i, \boldsymbol{p}_{-i} \right ) < l^{i} \left ( p_i', \boldsymbol{p}_{-i} \right ), \ \forall \boldsymbol{p}_{-i} \in \boldsymbol{\mathcal{P}}_{-i},
\end{equation}
and
\begin{equation}
    \label{eq_strict_dominance_MV_sufficient_second_moment}
    \mathrm{E} \left [ \Big ( L^{i} \left ( p_i, \boldsymbol{p}_{-i} \right ) \Big )^2 \right ] < \mathrm{E} \left [ \Big ( L^{i} \left ( p_i', \boldsymbol{p}_{-i} \right ) \Big )^2 \right ], \ \forall \boldsymbol{p}_{-i} \in \boldsymbol{\mathcal{P}}_{-i}.
\end{equation}

\end{definition}

Consider that path $p_i \in \mathcal{P}_i$ strictly dominates path $p_i' \in \mathcal{P}_i$ for player $i$ as defined in Definition \ref{def_strict_dominance_MV_new}; then, using Equation \eqref{eq_strict_dominance_MV_sufficient_mean}, for any $\boldsymbol{\sigma}_{-i} \in \boldsymbol{\Sigma}_{-i}$,
\begin{equation}
    \overline{l}^{i}(p_i, \boldsymbol{\sigma}_{-i}) = \sum_{\boldsymbol{p}_{-i} \in \boldsymbol{\mathcal{P}}_{-i}} \Big ( \boldsymbol{\sigma}(\boldsymbol{p}_{-i}) \cdot l^{i}(p_i, \boldsymbol{p}_{-i}) \Big ) < \sum_{\boldsymbol{p}_{-i} \in \boldsymbol{\mathcal{P}}_{-i}} \Big ( \boldsymbol{\sigma}(\boldsymbol{p}_{-i}) \cdot l^{i}(p_i', \boldsymbol{p}_{-i}) \Big ) = \overline{l}^{i}(p_i', \boldsymbol{\sigma}_{-i}).
\end{equation}
Note that $\overline{l}^{i}(p_i, \boldsymbol{\sigma}_{-i}) \leq \frac{\rho}{2}$ for all $p_i \in \mathcal{P}_i$, for all $\boldsymbol{\sigma}_{-i} \in \boldsymbol{\Sigma}_{-i}$, and for all $i \in [n]$ as a result of $l^{i}(\boldsymbol{p}) \leq \frac{\rho}{2}$ for all $\boldsymbol{p} \in \boldsymbol{\mathcal{P}}$ and for all $i \in [n]$. Hence, using the fact that the function $-f^2 + \rho \cdot f$ is increasing for $f \leq \frac{\rho}{2}$, for any $\boldsymbol{\sigma}_{-i} \in \boldsymbol{\Sigma}_{-i}$ we have
\begin{equation}
    \label{eq_intermediate_mean_strict_dominance_MV}
    - \left ( \overline{l}^{i}(p_i, \boldsymbol{\sigma}_{-i}) \right )^2 + \rho \cdot \overline{l}^{i}(p_i, \boldsymbol{\sigma}_{-i}) < - \left ( \overline{l}^{i}(p_i', \boldsymbol{\sigma}_{-i}) \right )^2 + \rho \cdot \overline{l}^{i}(p_i', \boldsymbol{\sigma}_{-i}).
\end{equation}
On the other hand, using Equation \eqref{eq_strict_dominance_MV_sufficient_second_moment}, we have
\begin{equation}
\label{eq_intermediate_second_moment_strict_dominance_MV}
    \begin{aligned}
    & \mathrm{E} \left [ \left ( \overline{L}^{i}(p_i, \boldsymbol{\sigma}_{-i}) \right)^2 \right ] = \sum_{\boldsymbol{p}_{-i} \in \boldsymbol{\mathcal{P}}_{-i}} \left ( \boldsymbol{\sigma}(\boldsymbol{p}_{-i}) \cdot \mathrm{E} \left [ \left ( L^{i}(p_i, \boldsymbol{p}_{-i}) \right)^2 \right ] \right ) \\
    < & \sum_{\boldsymbol{p}_{-i} \in \boldsymbol{\mathcal{P}}_{-i}} \left ( \boldsymbol{\sigma}(\boldsymbol{p}_{-i}) \cdot \mathrm{E} \left [ \left ( L^{i}(p_i', \boldsymbol{p}_{-i}) \right)^2 \right ] \right ) = \mathrm{E} \left [ \left ( \overline{L}^{i}(p_i', \boldsymbol{\sigma}_{-i}) \right)^2 \right ].
    \end{aligned}
\end{equation}
Finally, Equations \eqref{eq_intermediate_mean_strict_dominance_MV} and \eqref{eq_intermediate_second_moment_strict_dominance_MV} conclude that $\mathrm{Var} \left (\overline{L}^{i}(p_i, \boldsymbol{\sigma}_{-i}) \right ) + \rho \cdot \overline{l}^{i}(p_i, \boldsymbol{\sigma}_{-i}) < \mathrm{Var} \left (\overline{L}^{i}(p_i', \boldsymbol{\sigma}_{-i}) \right ) + \rho \cdot \overline{l}^{i}(p_i', \boldsymbol{\sigma}_{-i})$ for all $\boldsymbol{\sigma}_{-i} \in \boldsymbol{\Sigma}_{-i}$.

In order to find the mean-variance equilibrium for a stochastic congestion game, we use support enumeration.
For example, hypothesize $\boldsymbol{\mathcal{P}}' \coloneqq \{ \mathcal{P}_1', \mathcal{P}_2', \dots, \mathcal{P}_n' \}$ to be the support of a mean-variance equilibrium, where $\mathcal{P}_i'$ is the set of pure strategies of player $i$ that are played with non-zero probability and $\sigma_i(p_i)$ for $p_i \in \mathcal{P}_i'$ indicates the probability mass function on the support.
At equilibrium, player $i \in [n]$ should be indifferent between strategies in the set $\mathcal{P}_i'$, has no incentive to deviate to the rest of strategies in the set $\mathcal{P}_i \setminus \mathcal{P}_i'$, and the probability mass function over the support should add to one.
As a result, if there is a mean-variance equilibrium with the mentioned support, it is the solution of the following set of equations for $\boldsymbol{\sigma} \in \boldsymbol{\Sigma}$:
\begin{equation}
\label{eq_find_mixed_strategy_MV}
    \left\{
    \begin{array}{ll}
        \mathrm{Var} \left (\overline{L}^{i}(p_i', \boldsymbol{\sigma}_{-i}) \right ) + \rho \cdot \overline{l}^{i}(p_i', \boldsymbol{\sigma}_{-i}) \leq \mathrm{Var} \left (\overline{L}^{i}(p_i, \boldsymbol{\sigma}_{-i}) \right ) + \rho \cdot \overline{l}^{i}(p_i, \boldsymbol{\sigma}_{-i}), \forall p_i\in \mathcal{P}_i, p_i' \in \mathcal{P}_i', \forall i \in [n],\\
        \\
        \sum_{p_i \in \mathcal{P}_i'} \sigma_i(p_i) = 1, \forall i \in [n],\\
        \\
        \sigma_i(p_i) = 0, \forall p_i \in \mathcal{P}_i \setminus \mathcal{P}_i', \forall i \in [n].
    \end{array}
    \right.
\end{equation}
%%%mention the previous is polynomial in order n - 1 and the current one is polynomial in terms of 2(n - 1).

\subsection{CVaR$_\alpha$ Equilibrium}
\label{CVaR_sub_section}

The conditional value at risk level $\alpha$ (CVaR$_\alpha$) is another framework in statistics to measure risk and to address the risk-averse behavior.
%The CVaR$_\alpha$ method measures risk by the expected value of the random variable conditioned on its lower $\alpha$ quantile.
Applying this method to the proposed stochastic congestion game setting, the CVaR$_\alpha$ best response and CVaR$_\alpha$ equilibrium are defined below.

\begin{definition}
\label{def_best_response_mixed_CVaR_alpha}
Given the mixed strategy profile $\boldsymbol{\sigma}_{-i}$ of players $[n] \setminus i$, the set of mixed strategy CVaR$_\alpha$ best responses of player $i$ is the set of all probability distributions over the set
\begin{equation}
    \label{eq_mixed_best_response_CVaR_alpha}
    \underset{p_i \in \mathcal{P}_i}{\argmin} \ CVaR_\alpha \left (\overline{L}^{i}(p_i, \boldsymbol{\sigma}_{-i}) \right ) = \underset{p_i \in \mathcal{P}_i}{\argmin} \ \mathrm{E} \left [\overline{L}^{i}(p_i, \boldsymbol{\sigma}_{-i}) \Big | \overline{L}^{i}(p_i, \boldsymbol{\sigma}_{-i}) \geq v_\alpha^{i}(p_i, \boldsymbol{\sigma}_{-i}) \right ],
\end{equation}
where $v_\alpha^{i}(p_i, \boldsymbol{\sigma}_{-i})$ is a constant derived by solving the equality $P \left ( \overline{L}^{i}(p_i, \boldsymbol{\sigma}_{-i}) \geq v_\alpha^{i}(p_i, \boldsymbol{\sigma}_{-i}) \right ) = \alpha$ and the constant $0 < \alpha \leq 1$ is a hyper-parameter depicting the risk level.
Given the mixed strategy $\boldsymbol{\sigma}_{-i}$ of players $[n] \setminus i$, the CVaR$_\alpha$ best response set of player $i$'s strategies is denoted by $CB(\boldsymbol{\sigma}_{-i})$, which is in general a set-valued function.

\end{definition}

\begin{definition}
\label{def_mixed_strategy_CVaR_alpha}
A strategy profile $\boldsymbol{\sigma}^* = (\sigma_1^*, \sigma_2^*, \dots,$ $\sigma_N^*)$ is a CVaR$_\alpha$ equilibrium if and only if $\sigma_i^* \in CB(\boldsymbol{\sigma}_{-i}^*)$ for all $i \in [n]$.
\end{definition}

The existence of the CVaR$_\alpha$ equilibrium is discussed in the following theorem.

\begin{theorem}
\label{theorem_existence_CVaR_alpha}
For any finite $n$-player stochastic congestion game, a CVaR$_\alpha$ equilibrium exists.
\end{theorem}

The proof of Theorem \ref{theorem_existence_CVaR_alpha} is provided in Appendix \ref{proof_theorem_CVaR_alpha}.

The pure strategy CVaR$_\alpha$ best response of player $i$ to the pure strategy $\boldsymbol{p}_{-i}$ of players $[n] \setminus i$ is the set
\begin{equation}
    \label{pure_best_response_CVaR_alpha}
    \underset{p_i \in \mathcal{P}_i}{\argmin} \ CVaR_\alpha \left (L^{i}(p_i, \boldsymbol{p}_{-i}) \right ) = \underset{p_i \in \mathcal{P}_i}{\argmin} \ \mathrm{E} \left [L^{i}(p_i, \boldsymbol{p}_{-i}) \Big | L^{i}(p_i, \boldsymbol{p}_{-i}) \geq v_\alpha^{i}(p_i, \boldsymbol{p}_{-i}) \right ],
\end{equation}
where $v_\alpha^{i}(p_i, \boldsymbol{p}_{-i})$ is a constant derived by solving the equality $P \left ( L^{i}(p_i, \boldsymbol{p}_{-i}) \geq v_\alpha^{i}(p_i, \boldsymbol{p}_{-i}) \right ) = \alpha$ and the constant $0 < \alpha \leq 1$ is the hyper-parameter depicting risk level.
Given the pure strategy $\boldsymbol{p}_{-i}$ of players $[n] \setminus i$, the CVaR$_\alpha$ best response set of player $i$ in Equation \eqref{pure_best_response_CVaR_alpha} is denoted by $CB(\boldsymbol{p}_{-i})$ (overloading notation, $CB(.)$ is used for both pure and mixed strategy CVaR$_\alpha$ best responses).
As a result, a pure strategy profile $\boldsymbol{p}^* = (p_1^*, p_2^*, \dots, p_n^*)$ is a pure strategy CVaR$_\alpha$ equilibrium if and only if $p_i^* \in CB(\boldsymbol{p}_{-i}^*)$ for all $i \in [n]$.
A pure strategy $p_i \in \mathcal{P}_i$ of player $i$ strictly dominates a second pure strategy $p_i' \in \mathcal{P}_i$ of the player in pure strategy CVaR$_\alpha$ equilibrium if
\begin{equation}
    \label{eq_strict_dominance_CVaR_alpha_pure}
    \mathrm{E} \left [L^{i}(p_i, \boldsymbol{p}_{-i}) \Big | L^{i}(p_i, \boldsymbol{p}_{-i}) \geq v_\alpha^{i}(p_i, \boldsymbol{p}_{-i}) \right ] < \mathrm{E} \left [L^{i}(p_i', \boldsymbol{p}_{-i}) \Big | L^{i}(p_i', \boldsymbol{p}_{-i}) \geq v_\alpha^{i}(p_i', \boldsymbol{p}_{-i}) \right ], \ \forall \boldsymbol{p}_{-i} \in \boldsymbol{\mathcal{P}}_{-i},
\end{equation}
where $v_\alpha^{i}(p_i, \boldsymbol{p}_{-i})$ and $v_\alpha^{i}(p_i', \boldsymbol{p}_{-i})$ are constants derived by solving $P \left ( L^{i}(p_i, \boldsymbol{p}_{-i}) \geq v_\alpha^{i}(p_i, \boldsymbol{p}_{-i}) \right ) = \alpha$ and $P \left ( L^{i}(p_i', \boldsymbol{p}_{-i}) \geq v_\alpha^{i}(p_i', \boldsymbol{p}_{-i}) \right ) = \alpha$, and the constant $0 < \alpha \leq 1$ is the risk level hyper-parameter.
However, similar to the mean-variance equilibrium, strict dominance may not be derived from Equation \eqref{eq_strict_dominance_CVaR_alpha_pure} for mixed strategy CVaR$_\alpha$ equilibrium as described below.
%Consider path $p_i \in \mathcal{P}_i$ strictly dominates path $p_i' \in \mathcal{P}_i$ for player $i$ as defined in Definition \ref{def_strict_dominance_CVaR_alpha}.
Using Equation \eqref{dist_mixed} and $P \left ( \overline{L}^{i}(p_i, \boldsymbol{\sigma}_{-i}) \geq v_\alpha^{i}(p_i, \boldsymbol{\sigma}_{-i}) \right ) = \alpha$, the distribution of the random variable $\left ( \overline{L}^{i}(p_i, \boldsymbol{\sigma}_{-i}) \Big | \overline{L}^{i}(p_i, \boldsymbol{\sigma}_{-i}) \geq v_\alpha^{i}(p_i, \boldsymbol{\sigma}_{-i}) \right )$ is
\begin{equation}
\label{dist_mixed_CVaR_alpha}
\left ( \sum_{\boldsymbol{p}_{-i} \in \boldsymbol{\mathcal{P}}_{-i}} \Big ( f^{i}(x | (p_i, \boldsymbol{p}_{-i})) \cdot \boldsymbol{\sigma}(\boldsymbol{p}_{-i}) \Big ) \bigg / \alpha \right ) \cdot \mathbbm{1} \left \{ x \geq v_\alpha^{i}(p_i, \boldsymbol{\sigma}_{-i}) \right \}.
\end{equation}
As a result,
\begin{equation}
    \label{eq_strict_dominance_CVaR_alpha}
    \begin{aligned}
    & \mathrm{E} \left [\overline{L}^{i}(p_i, \boldsymbol{\sigma}_{-i}) \Big | \overline{L}^{i}(p_i, \boldsymbol{\sigma}_{-i}) \geq v_\alpha^{i}(p_i, \boldsymbol{\sigma}_{-i}) \right ] \\
    \overset{(a)}{=} & \ \frac{1}{\alpha} \cdot \sum_{\boldsymbol{p}_{-i} \in \boldsymbol{\mathcal{P}}_{-i}} \left ( \boldsymbol{\sigma}(\boldsymbol{p}_{-i}) \cdot \int_{-\infty}^\infty \Big ( x \cdot f^{i}(x | (p_i, \boldsymbol{p}_{-i})) \cdot \mathbbm{1} \left \{ x \geq v_\alpha^{i}(p_i, \boldsymbol{\sigma}_{-i}) \right \} \Big ) dx \right ) \\
    \overset{(b)}{=} & \ \frac{1}{\alpha} \cdot \sum_{\boldsymbol{p}_{-i} \in \boldsymbol{\mathcal{P}}_{-i}} \Bigg ( \boldsymbol{\sigma}(\boldsymbol{p}_{-i}) \cdot P \big ( L^{i}(p_i, \boldsymbol{p}_{-i}) \geq v_\alpha^{i}(p_i, \boldsymbol{\sigma}_{-i}) \big ) \times \\
    & \hspace{2.53cm}  \int_{v_\alpha^{i}(p_i, \boldsymbol{\sigma}_{-i})}^\infty \Big ( x \cdot \frac{f^{i}(x | (p_i, \boldsymbol{p}_{-i}))}{P \big ( L^{i}(p_i, \boldsymbol{p}_{-i}) \geq v_\alpha^{i}(p_i, \boldsymbol{\sigma}_{-i}) \big )} \Big ) dx \Bigg ) \\
    = & \ \frac{1}{\alpha} \cdot \sum_{\boldsymbol{p}_{-i} \in \boldsymbol{\mathcal{P}}_{-i}} \left ( \boldsymbol{\sigma}(\boldsymbol{p}_{-i}) \cdot P \big ( L^{i}(p_i, \boldsymbol{p}_{-i}) \geq v_\alpha^{i}(p_i, \boldsymbol{\sigma}_{-i}) \big ) \cdot \mathrm{E} \left [ L^{i}(p_i, \boldsymbol{p}_{-i}) \Big | L^{i}(p_i, \boldsymbol{p}_{-i}) \geq v_\alpha^{i}(p_i, \boldsymbol{\sigma}_{-i}) \right ] \right ),
    \end{aligned}
\end{equation}
where $(a)$ is true by using the pdf of the corresponding random variable in Equation \eqref{dist_mixed_CVaR_alpha} and switching the order of summation and integral and $(b)$ is true by multiplying and dividing by the term $P \big ( L^{i}(p_i, \boldsymbol{p}_{-i}) \geq v_\alpha^{i}(p_i, \boldsymbol{\sigma}_{-i}) \big )$.
As can be seen in Equation \eqref{eq_strict_dominance_CVaR_alpha},
%due to dependence of $v_\alpha^{p_i}(p_i, \boldsymbol{\sigma}_{-i})$ on $\boldsymbol{\sigma}_{-i}$,
it is not clear whether Equation \eqref{eq_strict_dominance_CVaR_alpha_pure} can result in $\mathrm{E} \left [\overline{L}^{i}(p_i, \boldsymbol{\sigma}_{-i}) \Big | \overline{L}^{i}(p_i, \boldsymbol{\sigma}_{-i}) \geq v_\alpha^{i}(p_i, \boldsymbol{\sigma}_{-i}) \right ] < \mathrm{E} \left [\overline{L}^{i}(p_i', \boldsymbol{\sigma}_{-i}) \Big | \overline{L}^{i}(p_i', \boldsymbol{\sigma}_{-i}) \geq v_\alpha^{i}(p_i', \boldsymbol{\sigma}_{-i}) \right ]$ for all $\boldsymbol{\sigma}_{-i} \in \boldsymbol{\Sigma}_{-i}$.
As a result, use of strict dominance in the mixed strategy CVaR$_\alpha$ equilibrium is not advised due to its complication.

In order to find the CVaR$_\alpha$ equilibrium for a stochastic congestion game, we use support enumeration.
For example, hypothesize $\boldsymbol{\mathcal{P}}' \coloneqq \{ \mathcal{P}_1', \mathcal{P}_2', \dots, \mathcal{P}_n' \}$ to be the support of a CVaR$_\alpha$ equilibrium, where $\mathcal{P}_i'$ is the set of pure strategies of player $i$ that are played with non-zero probability and $\sigma_i(p_i)$ for $p_i \in \mathcal{P}_i'$ indicates the probability mass function on the support.
At equilibrium, player $i \in [n]$ should be indifferent between strategies in the set $\mathcal{P}_i'$, has no incentive to deviate to the rest of strategies in the set $\mathcal{P}_i \setminus \mathcal{P}_i'$, and the probability mass function over the support should add to one.
As a result, if there is a CVaR$_\alpha$ equilibrium with the mentioned support, it is the solution of the following set of equations for $\boldsymbol{\sigma} \in \boldsymbol{\Sigma}$:
\begin{equation}
\label{eq_find_mixed_strategy_CVaR_alpha}
    \left\{
    \begin{array}{ll}
         \mathrm{E} \left [\overline{L}^{i}(p_i', \boldsymbol{\sigma}_{-i}) \Big | \overline{L}^{i}(p_i', \boldsymbol{\sigma}_{-i}) \geq v_\alpha^{i}(p_i', \boldsymbol{\sigma}_{-i}) \right ] \\
        \leq \mathrm{E} \left [\overline{L}^{i}(p_i, \boldsymbol{\sigma}_{-i}) \Big | \overline{L}^{i}(p_i, \boldsymbol{\sigma}_{-i}) \geq v_\alpha^{i}(p_i, \boldsymbol{\sigma}_{-i}) \right ], \forall p_i\in \mathcal{P}_i, p_i' \in \mathcal{P}_i', \forall i \in [n],\\
        \\
        \sum_{p_i \in \mathcal{P}_i'} \sigma_i(p_i) = 1, \forall i \in [n],\\
        \\
        \sigma_i(p_i) = 0, \forall p_i \in \mathcal{P}_i \setminus \mathcal{P}_i', \forall i \in [n].
    \end{array}
    \right.
\end{equation}

\begin{remark}
It is noteworthy that the polynomial terms in Equation \eqref{eq_find_mixed_strategy_RAE} for the risk-averse equilibrium are of degree $n - 1$ while the polynomial terms in Equation \eqref{eq_find_mixed_strategy_MV} for the mean-variance equilibrium are of degree $2 (n - 1)$ for $n$ number of players.
On the other hand, it is more complicated to solve for Equation \eqref{eq_find_mixed_strategy_CVaR_alpha} as the top $\alpha$ quantile of distributions should be calculated.
\end{remark}

\section{Numerical Results}
\label{numerical_results_congestion_game}
%The three models proposed here are not compared with the approach that maximizes the probability that the traveler arrives on time since those models do not consider the stochastisity of links into account, in other words, they do not consider stochastisity dependent to the load, they just consider a fixed distribution for the error no matter what the load is on the links of the network.
The risk-averse, mean-variance, and CVaR$_\alpha$ equilibria are numerically analyzed for Examples \ref{example1} and \ref{example2} in this section.
The price of anarchy for each of the mentioned equilibria is calculated as well.
In the end, extra examples are presented to shed light on the corner cases of each one of the equilibria and to provide insight on how to tackle such circumstances.

In order to find any of the three types of pure equilibria for the Pigou network in Example \ref{example1} with $n$ players, hypothesize that $m_1$ players choose link $1$ and $m_2 = n - m_1$ players choose link $2$ and check whether any players has any incentive in the corresponding sense of the equilibrium of the interest to change route, given the pure strategy of the other players.
If none of the players has any incentive to change route given the pure strategy of the rest of players, $(m_1, n - m_1)$ is a pure equilibrium, where $(m_1, m_2)$ denotes that $m_1$ players select link $1$ and $m_2$ players select link $2$.
By varying $m_1$ from zero to $n$ and taking the above procedure, the pure equilibrium is found if any exists.
Given a fixed number of players $m_1$ that choose link $1$, it is obvious that they all have the same incentive to change to link $2$ or stay in link $1$, and all of the $m_2 = n - m_1$ players have the same incentive to change to link $1$ or stay in link $2$.
As a result, if a specific player out of the $m_1$ players has no incentive to switch to link $2$ given the pure strategy of the other players, and a specific player out of the $m_2$ players has no incentive to switch to link $1$ given the pure strategy of the other players, $(m_1, m_2 = n - m_1)$ is a pure equilibrium.
In other words, $(m_1, m_2 = n - m_1)$ is a pure risk-averse equilibrium if
\begin{equation}
    \label{RAE_Pigou_numerical}
    \left\{
    \begin{array}{lll}
        P \big (L_1(m_1) \leq L_2(m_2 + 1) \big ) \geq 0.5,
        \\
        \\
        P \big (L_2(m_2) \leq L_1(m_1 + 1) \big ) \geq 0.5,
    \end{array}
    \right.
\end{equation}
where the first inequality is true since each player has two options, link $1$ and link $2$, so $P \big (L_1(m_1) \leq L_2(m_2 + 1) \big ) \geq P \big (L_2(m_2 + 1) \leq L_1(m_1) \big )$, and since random variables are continuous we have $P \big (L_1(m_1) \leq L_2(m_2 + 1) \big ) + P \big (L_2(m_2 + 1) \leq L_1(m_1) \big ) = 1$, which results in $P \big (L_1(m_1) \leq L_2(m_2 + 1) \big ) \geq 0.5$.
The second inequality is true due to a similar reasoning.
By varying $m_1$ from zero to $n$, if Equation \eqref{RAE_Pigou_numerical} holds for $(m_1, m_2 = n - m_1)$, it is a pure risk-averse equilibrium.

Similar to the above approach, $(m_1, m_2 = n - m_1)$ is a pure mean-variance equilibrium if
\begin{equation}
    \label{MV_Pigou_numerical}
    \left\{
    \begin{array}{lll}
        \mathrm{Var} \big (L_1(m_1) \big ) + \rho \cdot l_1(m_1) \leq \mathrm{Var} \big (L_2(m_2 + 1) \big ) + \rho \cdot l_2(m_2 + 1),
        \\
        \\
        \mathrm{Var} \big (L_2(m_2) \big ) + \rho \cdot l_2(m_2) \leq \mathrm{Var} \big (L_1(m_1 + 1) \big ) + \rho \cdot l_1(m_1 + 1).
    \end{array}
    \right.
\end{equation}
Again, by varying $m_1$ from zero to $n$, if Equation \eqref{MV_Pigou_numerical} holds for $(m_1, m_2 = n - m_1)$, it is a pure mean-variance equilibrium. Similarly, $(m_1, m_2 = n - m_1)$ is a pure CVaR$_\alpha$ equilibrium if
\begin{equation}
    \label{CVaR_Pigou_numerical}
    \left\{
    \begin{array}{lll}
        \mathrm{E} \big [ L_1(m_1) \big | L_1(m_1) \geq v_\alpha^1(m_1) \big ] \leq \mathrm{E} \big [ L_2(m_2 + 1) \big | L_2(m_2 + 1) \geq v_\alpha^2(m_2 + 1) \big ],
        \\
        \\
        \mathrm{E} \big [ L_2(m_2) \big | L_2(m_2) \geq v_\alpha^2(m_2) \big ] \leq \mathrm{E} \big [ L_1(m_1 + 1) \big | L_1(m_1 + 1) \geq v_\alpha^1(m_1 + 1) \big ],
    \end{array}
    \right.
\end{equation}
where $P \big ( L_1(m_1) \geq v_\alpha^1(m_1) \big ) = P \big ( L_2(m_2 + 1) \geq v_\alpha^2(m_2 + 1) \big ) = P \big ( L_2(m_2) \geq v_\alpha^2(m_2) \big ) = P \big ( L_1(m_1 + 1) \geq v_\alpha^1(m_1 + 1) \big ) = \alpha$.
By varying $m_1$ from zero to $n$, if Equation \eqref{CVaR_Pigou_numerical} holds for $(m_1, m_2 = n - m_1)$, it is a pure CVaR$_\alpha$ equilibrium.

\begin{figure}[t]
\centering
\includegraphics[width=0.9\textwidth]{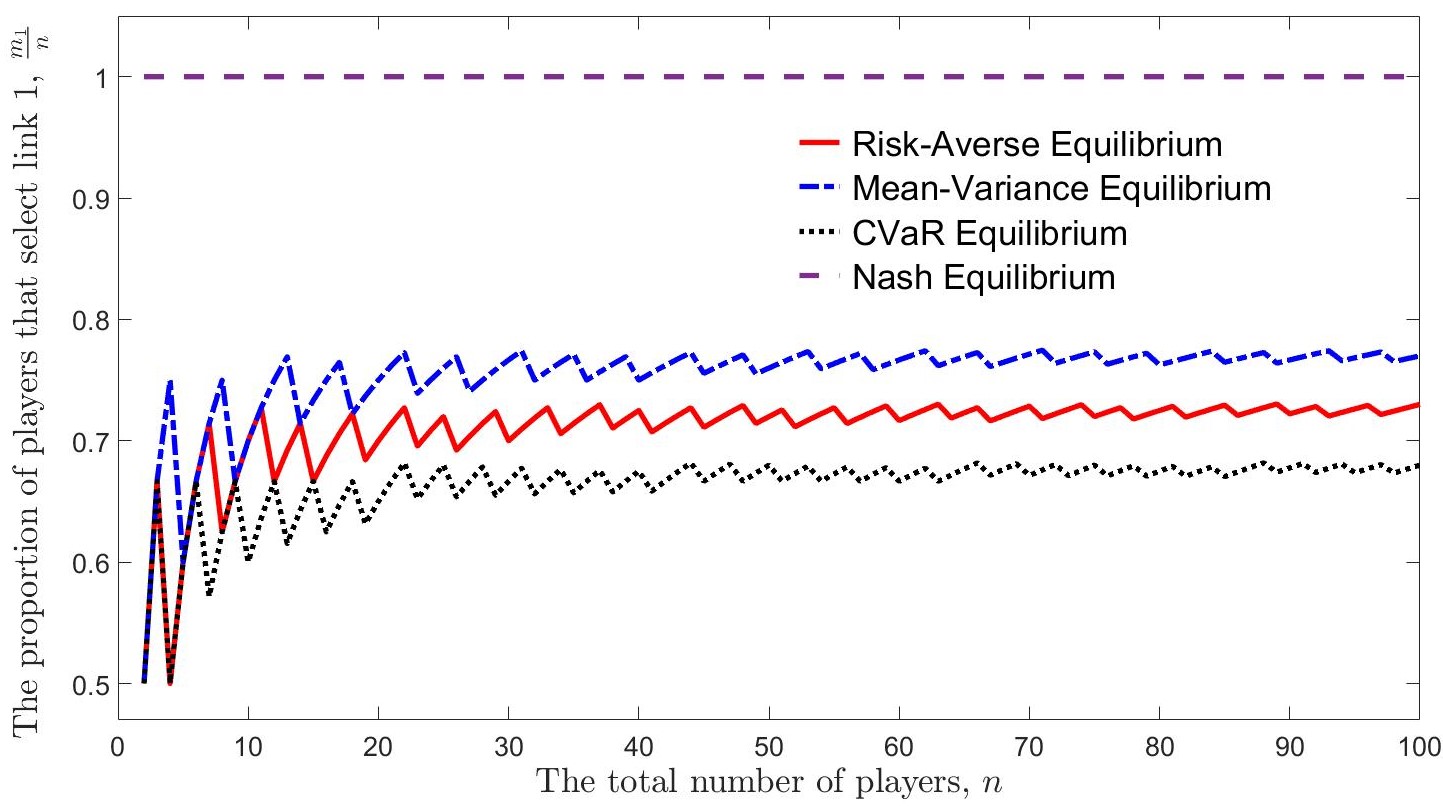}
\caption{The pure risk-averse, mean-variance ($\rho = 1$), CVaR$_\alpha$ ($\alpha = 0.1$), and Nash equilibria of the Pigou network in Example \ref{example1} are denoted for different numbers of players.}
\label{figure_Pigou_equilibrium}
\end{figure}

Note that the equilibrium in the Pigou network in Example \ref{example1} is characterized by $m_1$, since $m_2$ can be derived given $m_1$.
The pure risk-averse, mean-variance ($\rho = 1$), and CVaR$_\alpha$ ($\alpha = 0.1$) equilibria are found for the mentioned Pigou network and the proportion of players who select link $1$, i.e., $\frac{m_1}{n}$, is depicted in Figure \ref{figure_Pigou_equilibrium} for different values of $n$.
Under the Nash equilibrium, no matter what the probability distributions of latency over links look like, all players select link $1$ as it has less or equal latency in expectation.
Hence, $(n, 0)$ is the Nash equilibrium for all $n$, which corresponds to $\frac{m_1}{n} = 1$ as depicted in Figure \ref{figure_Pigou_equilibrium}.

\begin{figure}[t]
\centering
\includegraphics[width=0.9\textwidth]{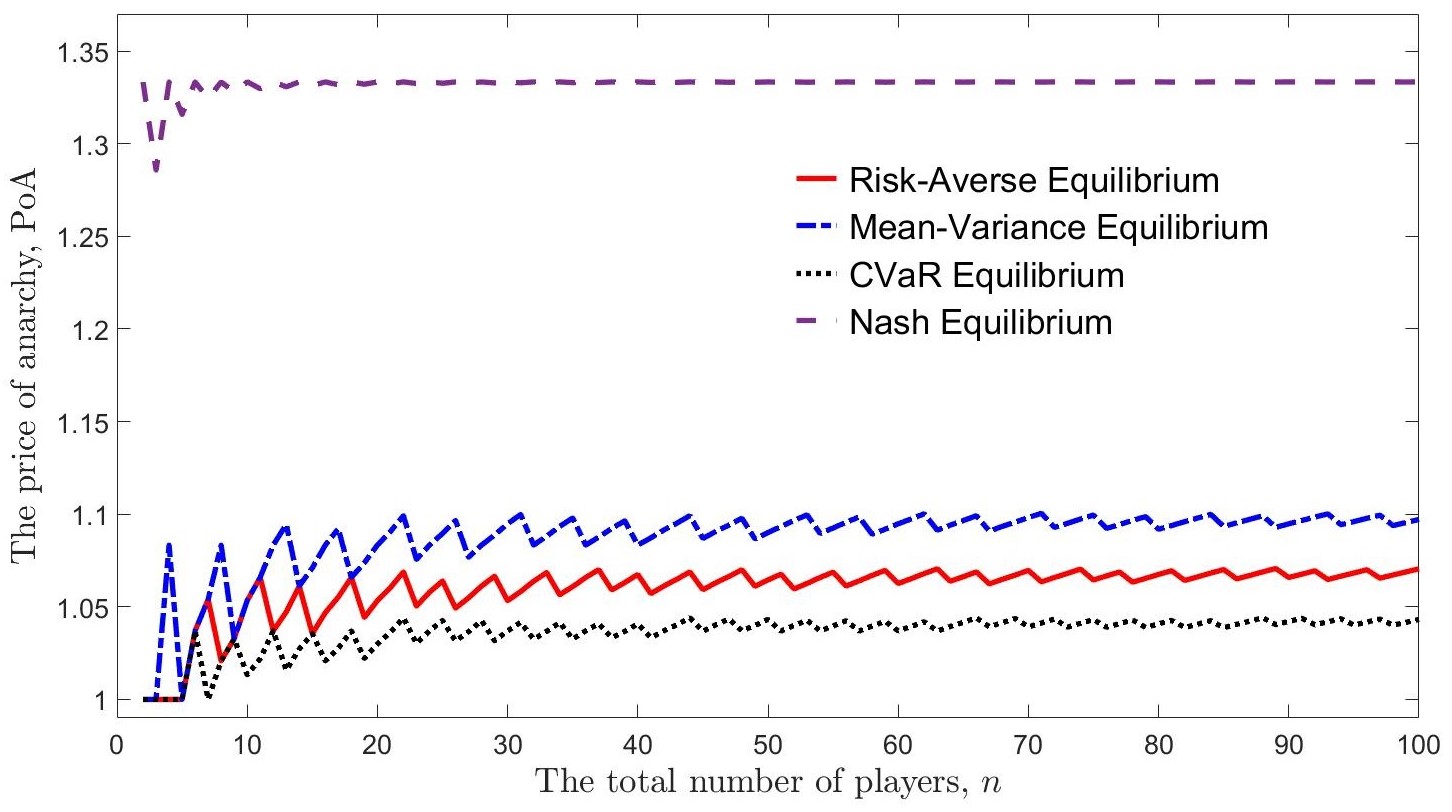}
\caption{The prices of anarchy for the risk-averse, mean-variance ($\rho = 1$), CVaR$_\alpha$ ($\alpha = 0.1$), and Nash equilibria of the Pigou network in Example \ref{example1} are plotted for different numbers of players.}
\label{figure_Pigou_PoA}
\end{figure}

The social delay/latency defined as the expected average delay/latency incurred by the $n$ players in the Pigou network in Example \ref{example1} under the pure strategy $(m_1, m_2)$ is $D(m_1) =  \frac{1}{n} \left ( m_1 \cdot \frac{m_1}{n} + (n - m_1) \right ) = \left ( \frac{m_1}{n} \right )^2 - \frac{m_1}{n} + 1$, which is minimized when $m_1 = \frac{n}{2}$ for an even $n$, and $m_1 = \lfloor \frac{n}{2} \rfloor$ and $m_1 = \lceil \frac{n}{2} \rceil$ for an odd $n$.
As a result, it is socially optimal that about half of the players take the top link and the rest take the bottom link to travel from source to destination in the Pigou network, which results in a social latency close to $\frac34$ for $n \gg 1$.
If players are risk-neutral and seek to minimize their expected latency given the strategy of the rest of players, which is how the Nash equilibrium models games, the social latency in the mentioned Pigou network equals to one for the Nash equilibrium $(n, 0)$.
In contrast, if players are risk-averse in the different senses discussed in this article, the social latency decreases compared to when players are risk-neutral; as a result, the price of anarchy decreases as depicted in Figure \ref{figure_Pigou_PoA}.
In this example, it is to the benefit of the society if players are risk-averse, which is the case as numerous studies in prospect theory discuss the fact that players in the real world often behave in a risk-averse manner.

Considering the Pigou network in a non-atomic setting, which corresponds to the case with infinite number of players, the socially optimal strategy is $(0.5, 0.5)$ with social latency of $\frac34$, where $(u_1, u_2)$ corresponds to $u_1$ fraction of players traveling along link $1$ and $u_2 = 1 - u_1$ fraction of players traveling along link $2$.
We numerically calculate that the risk-averse equilibrium is $(0.7303, 0.2697)$ with $\text{PoA} = 1.0707$, the mean-variance equilibrium with $\rho = 1$ is $(0.7750, 0.2250)$ with $\text{PoA} = 1.1008$, the CVaR$_\alpha$ equilibrium with $\alpha = 0.1$ is $(0.6822, 0.3178)$ with $\text{PoA} = 1.0442$, and the Nash equilibrium is $(1, 0)$ with $\text{PoA} = \frac43$.

In the Braess network in Example \ref{example2}, there are three paths from source to destination, $p_1 = (1, 2)$, $p_2 = (1, 5, 4)$, $p_3 = (3, 4)$, where links $SA, AD, SB, BD$, and $AB$ are denoted with $1, 2, 3, 4$, and $5$, respectively.
In order to find the three types of pure equilibria for the Braess network with $n$ players, hypothesize that $m^1$ players select path $p_1$, $m^2$ players select path $p_2$, and $n - m^1 - m^2$ players select path $p_3$, then check whether any players has any incentive in the corresponding sense of the equilibrium of the interest to change route, given the pure strategy of the other players.
If none of the players has any incentive to change route given the pure strategy of the rest of players, $(m^1, m^2, n - m^1 - m^2)$ is a pure equilibrium.
As a result, $(m^1, m^2, n - m^1 - m^2)$ is a pure risk-averse equilibrium if
\begin{equation}
    \label{RAE_Braess_numerical}
    \left\{
    \begin{array}{llllllll}
        P \big ( L^1 \leq \{ L^2, L^3 \} \big ) \geq \big \{ P \big ( L^2 \leq \{ L^1, L^3 \} \big ), P \big ( L^3 \leq \{ L^1, L^2 \} \big ) \big \}, \\
        \text{where } L^1 = L_1(m^1 + m^2) + L_2(m^1), L^2 = L_1(m^1 + m^2) + L_4(n - m^1 + 1), \text{and } \\
        L^3 = L_3(n - m^1 - m^2 + 1) + L_4(n - m^1 + 1), \\
        \\
        P \big ( L^2 \leq \{ L^1, L^3 \} \big ) \geq \big \{ P \big ( L^1 \leq \{ L^2, L^3 \} \big ), P \big ( L^3 \leq \{ L^1, L^2 \} \big ) \big \}, \\
        \text{where } L^1 = L_1(m^1 + m^2) + L_2(m^1 + 1), L^2 = L_1(m^1 + m^2) + L_4(n - m^1), \text{and } \\
        L^3 = L_3(n - m^1 - m^2 + 1) + L_4(n - m^1), \\
        \\
        P \big ( L^3 \leq \{ L^1, L^2 \} \big ) \geq \big \{ P \big ( L^1 \leq \{ L^2, L^3 \} \big ), P \big ( L^2 \leq \{ L^1, L^3 \} \big ) \big \}, \\
        \text{where } L^1 = L_1(m^1 + m^2 + 1) + L_2(m^1 + 1), L^2 = L_1(m^1 + m^2 + 1) + L_4(n - m^1), \text{and } \\
        L^3 = L_3(n - m^1 - m^2) + L_4(n - m^1).
    \end{array}
    \right.
\end{equation}
By varying $m^1$ from zero to $n$ and $m^2$ from $0$ to $n - m^1$, if Equation \eqref{RAE_Braess_numerical} holds for $(m^1, m^2, m^3 = n - m^1 - m^2)$, it is a pure risk-averse equilibrium.

Similar to the above approach, $(m^1, m^2, n - m^1 - m^2)$ is a pure mean-variance equilibrium if
\begin{equation}
    \label{MV_Braess_numerical}
    \left\{
    \begin{array}{llllllll}
        \mathrm{Var}(L^1) + \rho \cdot \mathrm{E}(L^1) \leq \big \{ \mathrm{Var}(L^2) + \rho \cdot \mathrm{E}(L^2), \mathrm{Var}(L^3) + \rho \cdot \mathrm{E}(L^3) \big \}, \\
        \text{where } L^1 = L_1(m^1 + m^2) + L_2(m^1), L^2 = L_1(m^1 + m^2) + L_4(n - m^1 + 1), \text{and } \\
        L^3 = L_3(n - m^1 - m^2 + 1) + L_4(n - m^1 + 1), \\
        \\
        \mathrm{Var}(L^2) + \rho \cdot \mathrm{E}(L^2) \leq \big \{ \mathrm{Var}(L^1) + \rho \cdot \mathrm{E}(L^1), \mathrm{Var}(L^3) + \rho \cdot \mathrm{E}(L^3) \big \}, \\
        \text{where } L^1 = L_1(m^1 + m^2) + L_2(m^1 + 1), L^2 = L_1(m^1 + m^2) + L_4(n - m^1), \text{and } \\
        L^3 = L_3(n - m^1 - m^2 + 1) + L_4(n - m^1), \\
        \\
        \mathrm{Var}(L^3) + \rho \cdot \mathrm{E}(L^3) \leq \big \{ \mathrm{Var}(L^1) + \rho \cdot \mathrm{E}(L^1), \mathrm{Var}(L^2) + \rho \cdot \mathrm{E}(L^2) \big \}, \\
        \text{where } L^1 = L_1(m^1 + m^2 + 1) + L_2(m^1 + 1), L^2 = L_1(m^1 + m^2 + 1) + L_4(n - m^1), \text{and } \\
        L^3 = L_3(n - m^1 - m^2) + L_4(n - m^1).
    \end{array}
    \right.
\end{equation}
By varying $m^1$ from zero to $n$ and $m^2$ from $0$ to $n - m^1$, if Equation \eqref{MV_Braess_numerical} holds for $(m^1, m^2, m^3 = n - m^1 - m^2)$, it is a pure risk-averse equilibrium.

Similar to the above approach, $(m^1, m^2, n - m^1 - m^2)$ is a pure CVaR$_\alpha$ equilibrium if
\begin{equation}
    \label{CVaR_Braess_numerical}
    \left\{
    \begin{array}{llllllll}
        \mathrm{E} \big [ L^1 \big | L^1 \geq v_\alpha^1 \big ] \leq \big \{ \mathrm{E} \big [ L^2 \big | L^2 \geq v_\alpha^2 \big ], \mathrm{E} \big [ L^3 \big | L^3 \geq v_\alpha^3 \big ] \big \}, \\
        \text{where } L^1 = L_1(m^1 + m^2) + L_2(m^1), L^2 = L_1(m^1 + m^2) + L_4(n - m^1 + 1), \\
        L^3 = L_3(n - m^1 - m^2 + 1) + L_4(n - m^1 + 1), \text{ and } P \big ( L^1 \geq v_\alpha^1 \big ) = P \big ( L^2 \geq v_\alpha^2 \big ) = P \big ( L^3 \geq v_\alpha^3 \big ) = \alpha \\
        \\
        \mathrm{E} \big [ L^2 \big | L^2 \geq v_\alpha^2 \big ] \leq \big \{ \mathrm{E} \big [ L^1 \big | L^1 \geq v_\alpha^1 \big ], \mathrm{E} \big [ L^3 \big | L^3 \geq v_\alpha^3 \big ] \big \}, \\
        \text{where } L^1 = L_1(m^1 + m^2) + L_2(m^1 + 1), L^2 = L_1(m^1 + m^2) + L_4(n - m^1), \\
        L^3 = L_3(n - m^1 - m^2 + 1) + L_4(n - m^1), \text{ and } P \big ( L^1 \geq v_\alpha^1 \big ) = P \big ( L^2 \geq v_\alpha^2 \big ) = P \big ( L^3 \geq v_\alpha^3 \big ) = \alpha\\
        \\
        \mathrm{E} \big [ L^3 \big | L^3 \geq v_\alpha^3 \big ] \leq \big \{ \mathrm{E} \big [ L^1 \big | L^1 \geq v_\alpha^1 \big ], \mathrm{E} \big [ L^2 \big | L^2 \geq v_\alpha^2 \big ] \big \}, \\
        \text{where } L^1 = L_1(m^1 + m^2 + 1) + L_2(m^1 + 1), L^2 = L_1(m^1 + m^2 + 1) + L_4(n - m^1), \\
        L^3 = L_3(n - m^1 - m^2) + L_4(n - m^1), \text{ and } P \big ( L^1 \geq v_\alpha^1 \big ) = P \big ( L^2 \geq v_\alpha^2 \big ) = P \big ( L^3 \geq v_\alpha^3 \big ) = \alpha.
    \end{array}
    \right.
\end{equation}
By varying $m^1$ from zero to $n$ and $m^2$ from $0$ to $n - m^1$, if Equation \eqref{CVaR_Braess_numerical} holds for $(m^1, m^2, m^3 = n - m^1 - m^2)$, it is a pure CVaR$_\alpha$ equilibrium.

\begin{figure}[t]
\centering
\includegraphics[width=0.9\textwidth]{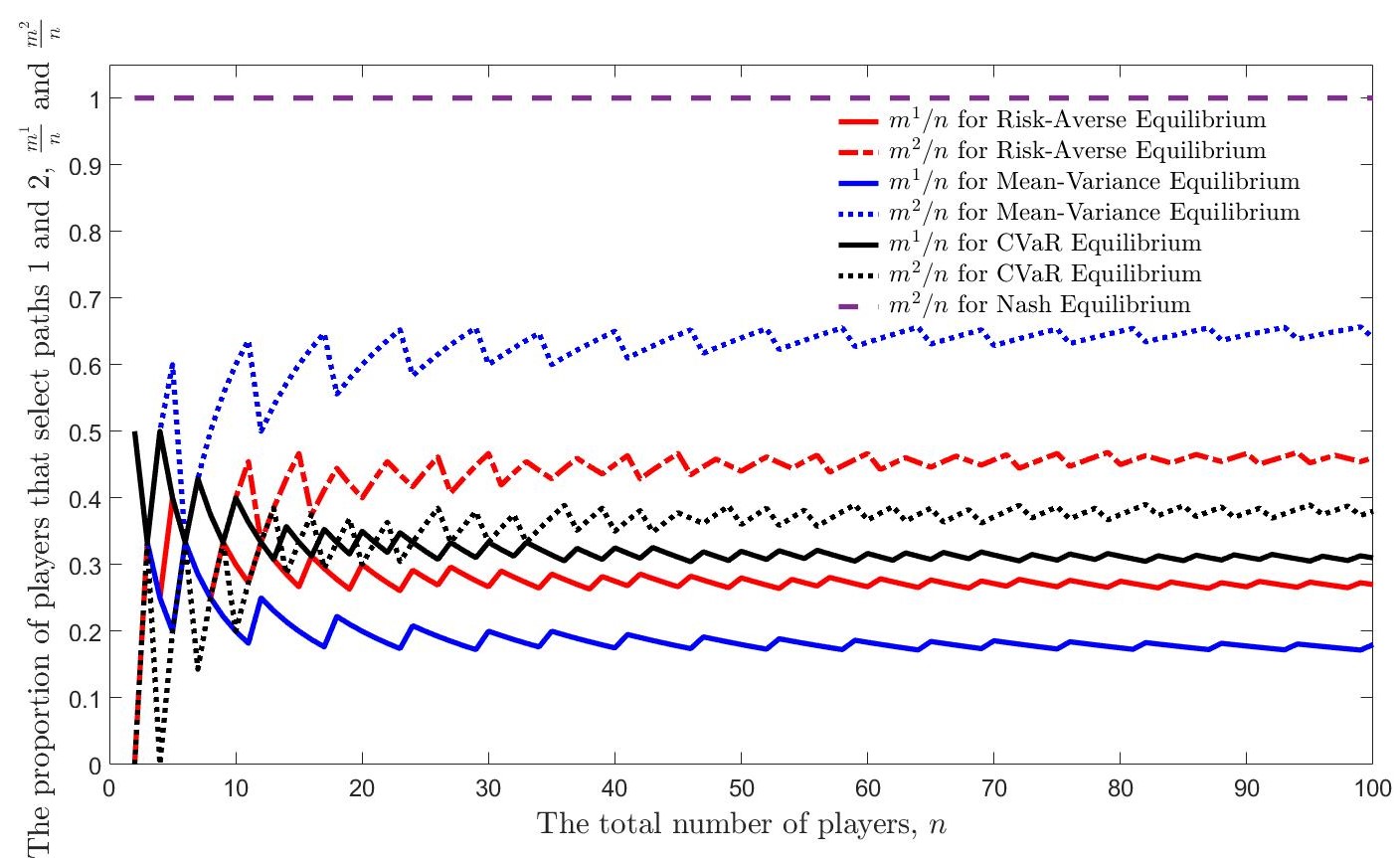}
\caption{The pure risk-averse, mean-variance ($\rho = 1$), CVaR$_\alpha$ ($\alpha = 0.1$), and Nash equilibria of the Braess network in Example \ref{example2} are denoted for different numbers of players.}
\label{figure_Braess_equilibrium}
\end{figure}

Note that the equilibrium in the Braess network in Example \ref{example2} is characterized by $m^1$ and $m^2$, since $m^3$ can be derived given $m^1$ and $m^2$.
The pure risk-averse, mean-variance ($\rho = 1$), and CVaR$_\alpha$ ($\alpha = 0.1$) equilibria are found for the mentioned Braess network and the proportions of players who select paths $1$ and $2$, i.e., $\frac{m^1}{n}$ and $\frac{m^2}{n}$, are depicted in Figure \ref{figure_Braess_equilibrium} for different values of $n$.
Under the Nash equilibrium, no matter what the probability distributions of latency over links look like, all players select path $2$ as it has less or equal latency in expectation.
Hence, $(0, n, 0)$ is the Nash equilibrium for all $n$, which corresponds to $\frac{m^2}{n} = 1$ and $\frac{m^1}{n} = \frac{m^3}{n} = 0$ as depicted in Figure \ref{figure_Braess_equilibrium}.

\begin{figure}[t]
\centering
\includegraphics[width=0.9\textwidth]{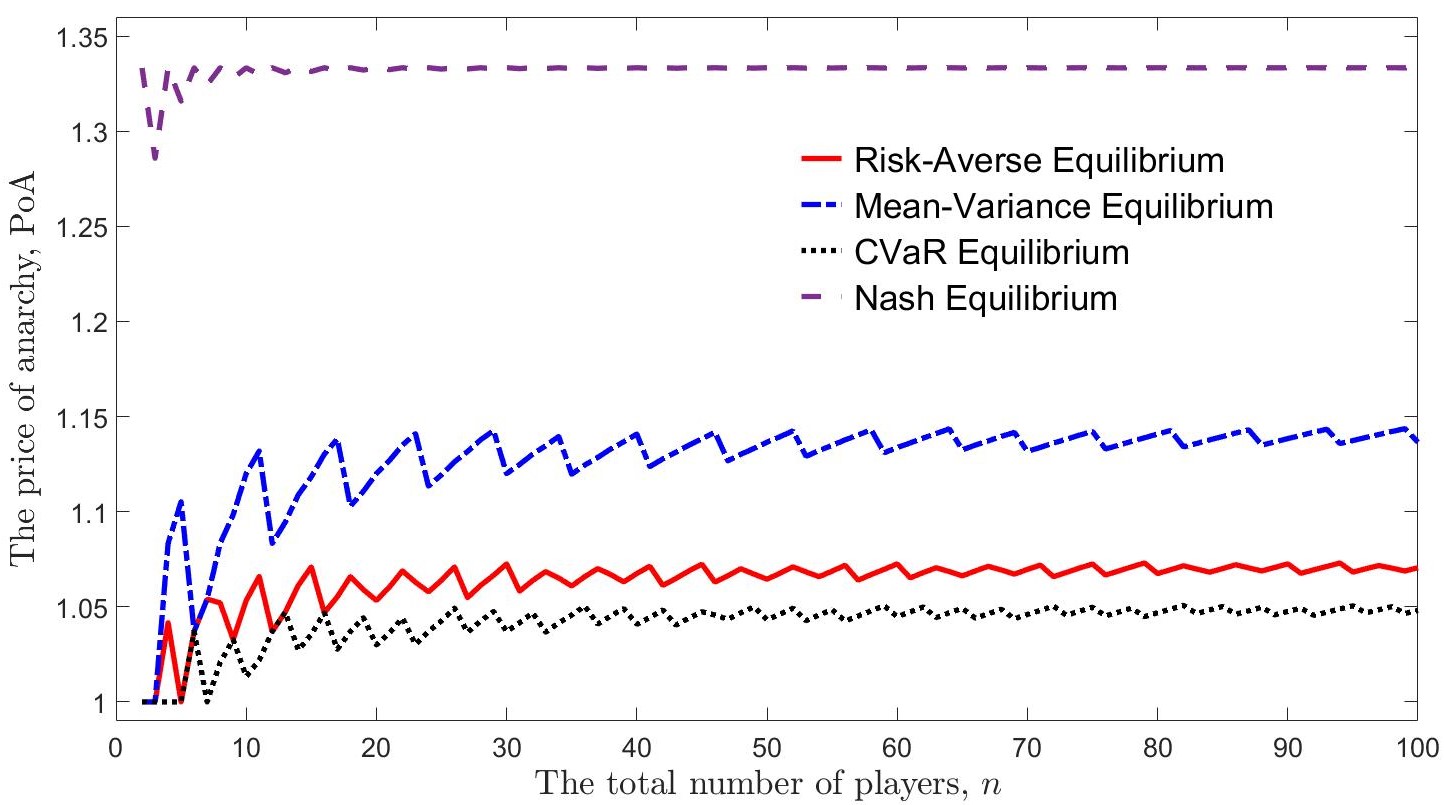}
\caption{The prices of anarchy for the risk-averse, mean-variance ($\rho = 1$), CVaR$_\alpha$ ($\alpha = 0.1$), and Nash equilibria of the Braess network in Example \ref{example2} are plotted for different numbers of players.}
\label{figure_Braess_PoA}
\end{figure}

The social delay/latency defined as the expected average delay/latency incurred by the $n$ players in the Braess network in Example \ref{example2} under the pure strategy $(m^1, m^2, m^3 = n - m^1 - m^2)$ is $D(m^1, m^2) =  \frac{1}{n} \cdot \left ( (m^1 + m^2) \cdot \frac{(m^1 + m^2)}{n} + m^1 + (n - m^1 - m^2) + (n - m^1) \cdot \frac{(n - m^1)}{n} \right ) = \frac{1}{n^2} \cdot \left ( 2 \left ( m^1 \right )^2 + \left ( m^2 \right )^2 + 2 m^1 m^2- 2 n m^1 - n m^2 + 2n^2 \right )$, which is minimized when $\left ( m^1 = \lfloor \frac{n}{2} \rfloor, m^2 = 0, m^3 = n - m^1 \right )$ or $\left ( m^1 = \lceil \frac{n}{2} \rceil, m^2 = 0, m^3 = n - m^1 \right )$.
As a result, it is socially optimal that about half of players take path $p_1$ and the rest take path $p_3$ to travel from source to destination in the Braess network, which results in a social latency close to $\frac32$ for $n \gg 1$.
If players are risk-neutral and seek to minimize their expected latency given the strategy of the rest of the players, which is how the Nash equilibrium models games, the social latency in the mentioned Braess network equals two for the Nash equilibrium $(0, n, 0)$.
In contrast, if players are risk-averse in the different senses discussed in this article, the social latency decreases compared to when players are risk-neutral; as a result, the price of anarchy decreases as depicted in Figure \ref{figure_Braess_PoA}.
In this example, it is again to the benefit of the society if players are risk-averse.

Considering the Braess network in a non-atomic setting, which corresponds to the case with infinite number of players, the socially optimal strategy is $(0.5, 0, 0.5)$ with social latency of $\frac32$, where $(u^1, u^2, u^3)$ corresponds to $u^1$ fraction of players travel along path $p_1$, $u^2$ fraction of players travel along path $p_2$, and $u^3 = 1 - u^1 - u^2$ fraction of players travel along path $p_3$.
We numerically calculate that the risk-averse equilibrium is $(0.2655, 0.4690, 0.2655)$ with $\text{PoA} = 1.0733$, the mean-variance equilibrium with $\rho = 1$ is $(0.1716, 0.6568, 0.1716)$ with $\text{PoA} = 1.1438$, the CVaR$_\alpha$ equilibrium with $\alpha = 0.1$ is $(0.3045, 0.3910, 0.3045)$ with $\text{PoA} = 1.0509$, and the Nash equilibrium is $(0, 1, 0)$ with $\text{PoA} = \frac43$.

\begin{figure}[t]
\centering
\includegraphics[width=1\textwidth]{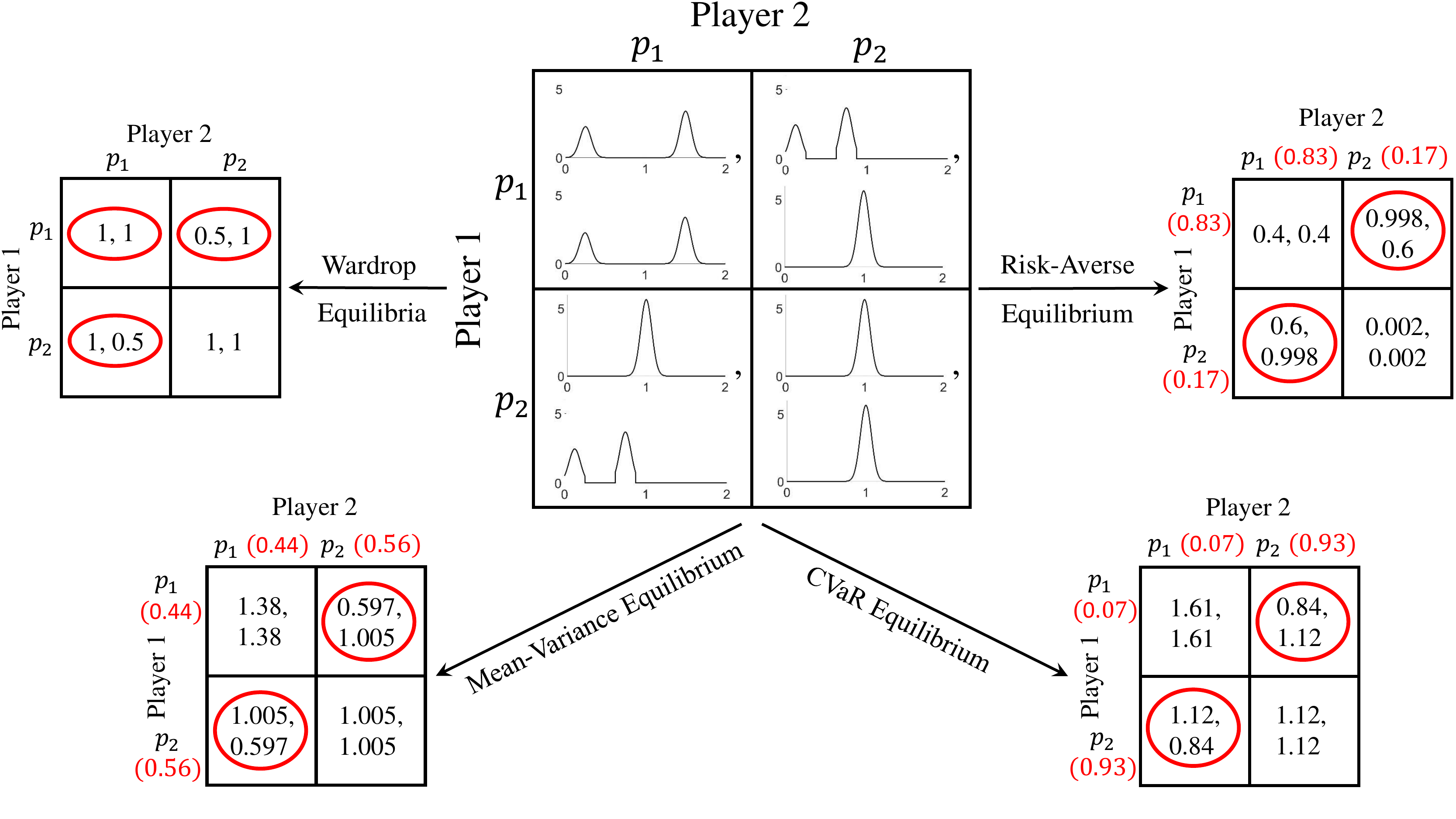}
\caption{The pure and mixed risk-averse, mean-variance ($\rho = 1$), CVaR$_\alpha$ ($\alpha = 0.1$), and Nash equilibria of the Pigou network in Example \ref{example1} for two players.}
\label{figure_Pigou_mixed_equilibria}
\end{figure}

Although it is more prevalent to use pure equilibrium for congestion games, we analyze the mixed equilibrium of the Pigou network in Example \ref{example1} for two players.
The underlying stochastic congestion game with the probability distributions of players' delays, the pure and mixed Nash, risk-averse, mean-variance, and CVaR equilibria are depicted in Figure \ref{figure_Pigou_mixed_equilibria}.
Recall that the (pure) price of anarchy of a congestion game is the maximum ratio $D(\boldsymbol{p}) \slash D(\boldsymbol{o})$ over all equilibria $\boldsymbol{p}$ of the game, where $\boldsymbol{o}$ is the socially optimum strategy.
As mentioned earlier, the optimum strategy for the Pigou network with two players is that one of the players travels along the top link and the other player travels along the bottom link which corresponds to the social delay of $\frac34$.
As a result, the (pure) price of anarchy for the Nash equilibria is $\frac43$.
On the other hand, the pure price of anarchy for the risk-averse, mean-variance, and CVaR equilibria is equal to one.
Furthermore, the price of anarchy among both pure and mixed equilibria for the risk-averse, mean-variance, and CVaR equilibria is $1.2405$, $1.1689$, and $1.2897$, respectively.

In the following, we present extra examples with the purpose of shedding light on drawbacks of the different equilibria in different scenarios and motivating more work to be done on a unified risk-averse framework.
Furthermore, the following examples suggest that careful consideration should be given to the choice of the equilibrium that best fits the application of the interest.

\subsection{Notes for Practitioners}
The intention of this subsection is to direct the attention of practitioners planning to implement risk-averse in-vehicle navigation to cases in which each of the proposed risk-averse equilibria may provide travelers with counterintuitive guidance.
To this end, three examples are discussed in the following to shed light on the implications of the three classes of risk-averse equilibria.
The examples are meant to be simple to convey the idea in a straightforward manner.

\begin{example}
\label{example3}
Consider a Pigou network with two parallel links, $1$ and $2$, between source and destination.
The travel times on links $1$ and $2$ are respectively independent random variables $L_1$ and $L_2$ with pdfs
\[
\begin{aligned}
    f_1(x) = \ & \alpha \bigg ( exp \left ( -100 \left (x - 14 \right )^2 \right ) \cdot \mathbbm{1} \left \{ 13 \leq x \leq 15 \right \}  + exp \left ( -100 \left (x - 19 \right )^2 \right ) \cdot \mathbbm{1} \left \{ 18 \leq x \leq 20 \right \}  \bigg ), \\
    f_2(x) = \ & \beta exp \left ( -100 \left (x - 20 \right )^2 \right ) \cdot \mathbbm{1} \left \{ 19 \leq x \leq 21 \right \},
\end{aligned}
\]
where $\alpha$ and $\beta$ are constants for which each of the two distributions integrate to one.
\end{example}

In Example \ref{example3}, the means and variances of travel times along links $1$ and $2$ are $l_1 = 16.5$, $\mathrm{Var}(L_1) = 6.255$, $l_2 = 20.0$, $\mathrm{Var}(L_2) = 0.005$, respectively, and $P(L_1 \leq L_2) = 1.0$.
As a result, although link $1$ has a higher variance than link $2$, not only is link $1$ shorter than link $2$ in expectation, but link $1$ is shorter than link $2$ almost certainly.
Hence, a rational traveler intends to take link $1$ for commute although its variance is higher than the variance of link $2$.
However, the mean-variance framework intends to keep a balance between lower expected travel time and lower uncertainty in travel time assuming that higher variance is against the spirit of risk-averse travelers.
In Example \ref{example3}, the mean-variance framework guides travelers to travel along link $2$ if $\rho < 1.7857$, which is not optimal from the perspective of a risk-averse traveler.
Note that both risk-averse equilibrium and CVaR$_\alpha$ equilibrium for any $\alpha \in [0, 1]$ guide travelers to traverse along link $1$ in this example.

\begin{example}
\label{example4}
Consider a Pigou network with two parallel links, $1$ and $2$, between source and destination.
The travel times on links $1$ and $2$ are respectively independent random variables $L_1$ and $L_2$ with pdfs
\[
\begin{aligned}
    f_1(x) = \ & \alpha \bigg ( 4exp \left ( -100 \left (x - 5 \right )^2 \right ) \cdot \mathbbm{1} \left \{ 4 \leq x \leq 6 \right \}  + exp \left ( -100 \left (x - 10 \right )^2 \right ) \cdot \mathbbm{1} \left \{ 9 \leq x \leq 11 \right \}  \bigg ), \\
    f_2(x) = \ & \beta \bigg ( 4exp \left ( -100 \left (x - 8 \right )^2 \right ) \cdot \mathbbm{1} \left \{ 7 \leq x \leq 9 \right \}  + exp \left ( -100 \left (x - 10 \right )^2 \right ) \cdot \mathbbm{1} \left \{ 9 \leq x \leq 11 \right \}  \bigg ),
\end{aligned}
\]
where $\alpha$ and $\beta$ are constants for which each of the two distributions integrate to one.
\end{example}

In Example \ref{example4}, the means and variances of travel times along links $1$ and $2$ are $l_1 = 6.0$, $\mathrm{Var}(L_1) = 4.005$, $l_2 = 8.4$, $\mathrm{Var}(L_2) = 0.645$, respectively, and $P(L_1 \leq L_2) = 0.82$.
Note that both distributions are the same over the interval $[9, 11]$; however, the traveler has a better opportunity of experiencing shorter travel time on the lower $0.8$ quantile of the distribution of link $1$ compared to that of link $2$.
Hence, a rational traveler intends to take link $1$ for commute although its variance is higher than the variance of link $2$.
Furthermore, $\mathrm{E} \left [ L_1 | L_1 \geq \alpha \right ] = \mathrm{E} \left [ L_2 | L_2 \geq \alpha \right ]$ for $\alpha \in [0, 0.2]$; hence, the CVaR$_\alpha$ framework is indifferent between the two links when $\alpha \in [0, 0.2]$, which can result in a counterintuitive route selection in Example \ref{example4}.
The mean-variance framework also guides travelers to traverse along link $2$ if $\rho < 1.4$, which is not optimal from the perspective of a risk-averse traveler.
Note that the risk-averse equilibrium guides travelers to traverse along link $1$ in this example as $P(L_1 \leq L_2) = 0.82$.

\begin{example}
\label{example5}
Consider a Pigou network with two parallel links, $1$ and $2$, between source and destination.
The travel times on links $1$ and $2$ are respectively independent random variables $L_1$ and $L_2$ with pdfs
\[
\begin{aligned}
    f_1(x) = \ & \beta exp \left ( -100 \left (x - 7 \right )^2 \right ) \cdot \mathbbm{1} \left \{ 6 \leq x \leq 8 \right \}, \\
    f_2(x) = \ & \alpha \bigg ( 7exp \left ( -100 \left (x - 5 \right )^2 \right ) \cdot \mathbbm{1} \left \{ 4 \leq x \leq 6 \right \}  + 3exp \left ( -100 \left (x - 10 \right )^2 \right ) \cdot \mathbbm{1} \left \{ 9 \leq x \leq 11 \right \}  \bigg ),
\end{aligned}
\]
where $\alpha$ and $\beta$ are constants for which each of the two distributions integrate to one.
\end{example}

In Example \ref{example5}, the means and variances of travel times along links $1$ and $2$ are $l_1 = 7.0$, $\mathrm{Var}(L_1) = 0.005$, $l_2 = 6.5$, $\mathrm{Var}(L_2) = 5.255$, respectively, and $P(L_2 \leq L_1) = 0.7$.
Although the expected travel time along link $2$ is less than that along link $1$ and it is more likely that the travel time along link $2$ is shorter than travel time along link $1$, the travel time along link $2$ is concentrated around $10$ with probability $0.3$ which is somewhat larger than the concentration of travel time around $7$ when traveling along link $1$.
Hence, a risk-averse traveler may prefer to take link $1$ for commute although its expected travel time is higher than the expected travel time of link $2$ to avoid a long travel time.
However, the risk-averse equilibrium guides travelers to traverse along link $2$, which may not be optimal from the perspective of a risk-averse traveler.
Note that the CVaR$_\alpha$ equilibrium for $\alpha < 0.748$ and mean-variance equilibrium for $\rho < 10.5$ guide travelers to traverse along link $1$ in this example.

\section{Conclusion and Future Work}
\label{conclusion_future_congestion_game}
A stochastic atomic congestion game with incomplete information on travel times along arcs of a traffic/telecommunication network is studied in this work from a risk-averse perspective.
Risk-averse travelers intend to make decisions based on probability statements regarding their travel options rather than simply taking the average travel time into account.
In order to put this into perspective, we propose three classes of equilibria, i.e., risk-averse equilibrium (RAE), mean-variance equilibrium (MVE), and CVaR$_\alpha$ equilibrium (CVaR$_\alpha$E).
The MV and CVaR$_\alpha$ equilibria are studied in the literature for networks with simplifying assumptions such as that the probability distributions of link delays are load independent or link delays are independent, which are not the case in this article.
The notions of best responses in risk-averse, mean-variance, and CVaR$_\alpha$ equilibria are based on maximizing the probability of traveling along the shortest path, minimizing a linear combination of mean and variance of path delay, and minimizing the expected delay at a specified risky quantile of the delay distributions, respectively.
We prove that the risk-averse, mean-variance, and CVaR$_\alpha$ equilibria exist for any finite stochastic atomic congestion game.
Although proving bounds on the price of anarchy (PoA) is not the focus of this work, we numerically study the impact of risk-averse equilibria on PoA and observe that the Braess paradox may not occur to the extent presented originally and the PoA may improve upon using any of the proposed equilibria in both Braess and Pigou networks.
Promising future directions are to study non-atomic, instead of atomic, stochastic congestion games in the proposed three classes of equilibria in their general case where the arc delay distributions are load dependent and not necessarily independent of each other, to find bounds on the price of anarchy for the proposed three classes of equilibria, and to find a unified class of equilibrium that captures risk-aversion for a broader class of travel time distributions in traffic/telecommunication networks.

\bibliographystyle{informs2014trsc}
\bibliography{myrefs}

\begin{appendices}
\label{appendix}

\section{Proof of Theorem \ref{theorem_existence_RAE}}
\label{proof_theorem_RAE}
Let $\boldsymbol{RB}: \boldsymbol{\Sigma} \rightarrow \boldsymbol{\Sigma}$ be the risk-averse best response function where $\boldsymbol{RB}(\boldsymbol{\sigma}) = \big (RB(\boldsymbol{\sigma}_{-1}), RB(\boldsymbol{\sigma}_{-2}),$ $\dots, RB(\boldsymbol{\sigma}_{-N}) \big )$.
It is easy to see that the existence of a fixed point $\boldsymbol{\sigma}^* \in \boldsymbol{\Sigma}$ for the risk-averse best response function, i.e., $\boldsymbol{\sigma}^* \in \boldsymbol{RB}(\boldsymbol{\sigma}^*)$, proves the existence of a risk-averse equilibrium.
The following four conditions of the Kakutani’s Fixed Point Theorem are shown to be satisfied for the function $\boldsymbol{RB}(\boldsymbol{\sigma})$ to prove the existence of a fixed point for the function.
\begin{enumerate}[leftmargin=*]
    \item The domain of function $\boldsymbol{RB}(.)$ is a non-empty, compact, and convex subset of a finite dimensional Euclidean space:
    $\boldsymbol{\Sigma}$ is the Cartesian product of non-empty simplices as each player has at least one strategy to play; furthermore, each of the elements of $\boldsymbol{\Sigma}$ is between zero and one, so $\boldsymbol{\Sigma}$ is non-empty, convex, bounded, and closed containing all its limit points.
    %Moreover, if $\boldsymbol{\sigma}_1 \in \boldsymbol{\Sigma}$ and $\boldsymbol{\sigma}_2 \in \boldsymbol{\Sigma}$, 
    \item $\boldsymbol{RB(\boldsymbol{\sigma})} \neq \oldemptyset$, $\forall \boldsymbol{\sigma} \in \boldsymbol{\Sigma}$:
    The set in Equation \eqref{eq_mixed_best_response} is non-empty as maximum exists over a finite number of values.
    As a result, $RB(\boldsymbol{\sigma}_{-i})$ is non-empty for all $i \in [n]$ since it is the set of all probability distributions over the corresponding mentioned non-empty set.
    \item The co-domain of function $\boldsymbol{RB}(.)$ is a convex set for all $\boldsymbol{\sigma} \in \boldsymbol{\Sigma}$:
    It suffices to prove that $RB(\boldsymbol{\sigma}_{-i})$ is a convex set for all $\boldsymbol{\sigma}_{-i} \in \boldsymbol{\Sigma}_{-i}$ and for all $i \in [n]$.
    %$RB(\boldsymbol{\sigma}_{-i})$ is a convex set for all $\boldsymbol{\sigma}_{-i} \in \boldsymbol{\Sigma}_{-i}$:
        For any $i \in [n]$, if $\sigma_i, \sigma_i' \in RB(\boldsymbol{\sigma}_{-i})$, we need to prove that $\lambda \sigma_i + (1 - \lambda) \sigma_i' \in RB(\boldsymbol{\sigma}_{-i})$ for any $\lambda \in [0, 1]$ and for any $\boldsymbol{\sigma}_{-i} \in \boldsymbol{\Sigma}_{-i}$.
        Let the supports of $\sigma_i$ and $\sigma_i'$ be defined as $supp(\sigma_i) = \{p_i \in \mathcal{P}_i: \sigma_i(p_i) > 0\}$ and $supp(\sigma_i') = \{p_i \in \mathcal{P}_i: \sigma_i'(p_i) > 0\}$, respectively.
        It is concluded from the definition of the risk-averse best response in Definition \ref{def_best_response_mixed_RAE} that $supp(\sigma_i), supp(\sigma_i') \subseteq \underset{p_i \in \mathcal{P}_i}{\argmax} \ P \left ( \overline{L}^{i}(p_i, \boldsymbol{\sigma}_{-i}) \leq \overline{\boldsymbol{L}}^{i}(\mathcal{P}_i \setminus p_i, \boldsymbol{\sigma}_{-i}) \right )$, which results in $supp(\sigma_i) \cup supp(\sigma_i') \subseteq \underset{p_i \in \mathcal{P}_i}{\argmax} \ P \left ( \overline{L}^{i}(p_i, \boldsymbol{\sigma}_{-i}) \leq \overline{\boldsymbol{L}}^{i}(\mathcal{P}_i \setminus p_i, \boldsymbol{\sigma}_{-i}) \right )$.
        As a result, using the definition of risk-averse best response, any probability distribution over the set $supp(\sigma_i) \cup supp(\sigma_i')$ is a risk-averse best response to $\boldsymbol{\sigma}_{-i}$.
        It is trivial that the mixed strategy $\lambda \sigma_i + (1 - \lambda) \sigma_i'$ is a valid probability distribution over the set $supp(\sigma_i) \cup supp(\sigma_i')$ for any $\lambda \in [0, 1]$, so $\lambda \sigma_i + (1 - \lambda) \sigma_i' \in RB(\boldsymbol{\sigma}_{-i})$ for any $\lambda \in [0, 1]$ and for any $\boldsymbol{\sigma}_{-i} \in \boldsymbol{\Sigma}_{-i}$ that completes the convexity proof of the set $RB(\boldsymbol{\sigma}_{-i})$.
    \item $\boldsymbol{RB}(\boldsymbol{\sigma})$ has a closed graph:
    %The fact that $\boldsymbol{RB}(\boldsymbol{\sigma})$ has a closed graph is proved by contradiction. %$\boldsymbol{RB}(\boldsymbol{\sigma})$ has a closed graph:
        $\boldsymbol{RB}(\boldsymbol{\sigma})$ has a closed graph if for any sequence $\{\boldsymbol{\sigma}^m, \widehat{\boldsymbol{\sigma}}^m\} \rightarrow \{\boldsymbol{\sigma}, \widehat{\boldsymbol{\sigma}}\}$ with $\widehat{\boldsymbol{\sigma}}^m \in \boldsymbol{RB}(\boldsymbol{\sigma}^m)$ for all $m \in \mathbb{N}$, we have $\widehat{\boldsymbol{\sigma}} \in \boldsymbol{RB}(\boldsymbol{\sigma})$.
        Proof by contradiction is used to show that $\boldsymbol{RB}(\boldsymbol{\sigma})$ has a closed graph.
        %Proof by contradiction is used to show that $\boldsymbol{RB}(\boldsymbol{\sigma})$ has a closed graph.
        Consider by contradiction that $\boldsymbol{RB}(\boldsymbol{\sigma})$ does not have a closed graph, so there exists a sequence $\{\boldsymbol{\sigma}^m, \widehat{\boldsymbol{\sigma}}^m\} \rightarrow \{\boldsymbol{\sigma}, \widehat{\boldsymbol{\sigma}}\}$ with $\widehat{\boldsymbol{\sigma}}^m \in \boldsymbol{RB}(\boldsymbol{\sigma}^m)$ for all $m \in \mathbb{N}$, but $\widehat{\boldsymbol{\sigma}} \notin \boldsymbol{RB}(\boldsymbol{\sigma})$.
        As a result, there exists some $i \in [n]$ such that $\widehat{\sigma}_i \notin RB(\boldsymbol{\sigma}_{-i})$.
        Using the definition of risk-averse best response in Definition \ref{def_best_response_mixed_RAE}, there exists $p_i' \in supp(RB(\boldsymbol{\sigma}_{-i}))$, $\widehat{p}_i \in supp(\widehat{\sigma}_i)$, and some $\epsilon > 0$ such that
        \begin{equation}
            \label{eq_contradiction_1}
            P \left ( \overline{L}^{i}(p_i', \boldsymbol{\sigma}_{-i}) \leq \overline{\boldsymbol{L}}^{i}(\mathcal{P}_i \setminus p_i', \boldsymbol{\sigma}_{-i}) \right ) > P \left ( \overline{L}^{i}(\widehat{p}_i, \boldsymbol{\sigma}_{-i}) \leq \overline{\boldsymbol{L}}^{i}(\mathcal{P}_i \setminus \widehat{p}_i, \boldsymbol{\sigma}_{-i}) \right ) + 3 \epsilon.
        \end{equation}
        Since the latencies over edges are continuous random variables and $\boldsymbol{\sigma}_{-i}^m \rightarrow \boldsymbol{\sigma}_{-i}$, for any $\epsilon > 0$, there exists a sufficiently large $m_1$ such that we have the following for $m \geq m_1$:
        \begin{equation}
            \label{eq_contradiction_2}
            P \left ( \overline{L}^{i}(p_i', \boldsymbol{\sigma}_{-i}^m) \leq \overline{\boldsymbol{L}}^{i}(\mathcal{P}_i \setminus p_i', \boldsymbol{\sigma}_{-i}^m) \right ) > P \left ( \overline{L}^{i}(p_i', \boldsymbol{\sigma}_{-i}) \leq \overline{\boldsymbol{L}}^{i}(\mathcal{P}_i \setminus p_i', \boldsymbol{\sigma}_{-i}) \right ) - \epsilon.
        \end{equation}
        By adding inequalities with the same direction in Equations \eqref{eq_contradiction_1} and \eqref{eq_contradiction_2}, for $m \geq m_1$ we have 
        \begin{equation}
            \label{eq_contradiction_3}
            P \left ( \overline{L}^{i}(p_i', \boldsymbol{\sigma}_{-i}^m) \leq \overline{\boldsymbol{L}}^{i}(\mathcal{P}_i \setminus p_i', \boldsymbol{\sigma}_{-i}^m) \right ) > P \left ( \overline{L}^{i}(\widehat{p}_i, \boldsymbol{\sigma}_{-i}) \leq \overline{\boldsymbol{L}}^{i}(\mathcal{P}_i \setminus \widehat{p}_i, \boldsymbol{\sigma}_{-i}) \right ) + 2 \epsilon.
        \end{equation}
        For the same reason as of Equation \eqref{eq_contradiction_2}, for any $\epsilon > 0$, there exists a sufficiently large $m_2$ such that we have the following for $m \geq m_2$:
        \begin{equation}
            \label{eq_contradiction_4}
            P \left ( \overline{L}^{i}(\widehat{p}_i, \boldsymbol{\sigma}_{-i}) \leq \overline{\boldsymbol{L}}^{i}(\mathcal{P}_i \setminus \widehat{p}_i, \boldsymbol{\sigma}_{-i}) \right ) > P \left ( \overline{L}^{i}(\widehat{p}_i^m, \boldsymbol{\sigma}_{-i}^m) \leq \overline{\boldsymbol{L}}^{i}(\mathcal{P}_i \setminus \widehat{p}_i^m, \boldsymbol{\sigma}_{-i}^m) \right ) - \epsilon,
        \end{equation}
        where $\widehat{p}_i^m \in supp(RB(\boldsymbol{\sigma}_{-i}^m))$.
        By adding the inequalities with the same direction in Equations \eqref{eq_contradiction_3} and \eqref{eq_contradiction_4}, for $m \geq \max \{m_1, m_2\}$ we have
        \begin{equation}
            \label{eq_contradiction_5}
            P \left ( \overline{L}^{i}(p_i', \boldsymbol{\sigma}_{-i}^m) \leq \overline{\boldsymbol{L}}^{i}(\mathcal{P}_i \setminus p_i', \boldsymbol{\sigma}_{-i}^m) \right ) > P \left ( \overline{L}^{i}(\widehat{p}_i^m, \boldsymbol{\sigma}_{-i}^m) \leq \overline{\boldsymbol{L}}^{i}(\mathcal{P}_i \setminus \widehat{p}_i^m, \boldsymbol{\sigma}_{-i}^m) \right ) + \epsilon.
        \end{equation}
        Equation \eqref{eq_contradiction_5} contradicts the fact that $\widehat{p}_i^m \in supp(RB(\boldsymbol{\sigma}_{-i}^m))$, which completes the proof that $\boldsymbol{RB}(\boldsymbol{\sigma})$ has a closed graph.
\end{enumerate}
As listed above, the risk-averse best response function $\boldsymbol{RB}(\boldsymbol{\sigma})$ satisfies the four conditions of Kakutani's Fixed Point Theorem.
As a direct result, for any finite $n$-player stochastic congestion game, there exists $\boldsymbol{\sigma}^* \in \boldsymbol{\Sigma}$ such that $\boldsymbol{\sigma}^* \in \boldsymbol{RB}(\boldsymbol{\sigma}^*)$, which completes the existence proof of a risk-averse equilibrium for such games.
\Halmos

\section{Proof of Theorem \ref{theorem_existence_MV}}
\label{proof_theorem_MV}
Let $\boldsymbol{MB}: \boldsymbol{\Sigma} \rightarrow \boldsymbol{\Sigma}$ be the mean-variance best response function where $\boldsymbol{MB}(\boldsymbol{\sigma}) = \big (MB(\boldsymbol{\sigma}_{-1}), MB(\boldsymbol{\sigma}_{-2}),$ $\dots, MB(\boldsymbol{\sigma}_{-N}) \big )$.
It is easy to see that the existence of a fixed point $\boldsymbol{\sigma}^* \in \boldsymbol{\Sigma}$ for the mean-variance best response function, i.e., $\boldsymbol{\sigma}^* \in \boldsymbol{MB}(\boldsymbol{\sigma}^*)$, proves the existence of a mean-variance equilibrium.
The following four conditions of the Kakutani’s Fixed Point Theorem are shown to be satisfied for the function $\boldsymbol{MB}(\boldsymbol{\sigma})$ to prove the existence of a fixed point for the function.
\begin{enumerate}[leftmargin=*]
    \item The domain of function $\boldsymbol{MB}(.)$ is a non-empty, compact, and convex subset of a finite dimensional Euclidean space:
    $\boldsymbol{\Sigma}$ is the Cartesian product of non-empty simplices as each player has at least one strategy to play; furthermore, each of the elements of $\boldsymbol{\Sigma}$ is between zero and one, so $\boldsymbol{\Sigma}$ is non-empty, convex, bounded, and closed containing all its limit points.
    %Moreover, if $\boldsymbol{\sigma}_1 \in \boldsymbol{\Sigma}$ and $\boldsymbol{\sigma}_2 \in \boldsymbol{\Sigma}$, 
    \item $\boldsymbol{MB(\boldsymbol{\sigma})} \neq \oldemptyset$, $\forall \boldsymbol{\sigma} \in \boldsymbol{\Sigma}$:
    The set in Equation \eqref{eq_mixed_best_response_MV} is non-empty as minimum exists over a finite number of values.
    As a result, $MB(\boldsymbol{\sigma}_{-i})$ is non-empty for all $i \in [n]$ since it is the set of all probability distributions over the corresponding mentioned non-empty set.
    \item The co-domain of function $\boldsymbol{MB}(.)$ is a convex set for all $\boldsymbol{\sigma} \in \boldsymbol{\Sigma}$:
    It suffices to prove that $MB(\boldsymbol{\sigma}_{-i})$ is a convex set for all $\boldsymbol{\sigma}_{-i} \in \boldsymbol{\Sigma}_{-i}$ and for all $i \in [n]$.
    %Each of the above three cases is proved separately below.
    %$MB(\boldsymbol{\sigma}_{-i})$ is a convex set for all $\boldsymbol{\sigma}_{-i} \in \boldsymbol{\Sigma}_{-i}$:
        For any $i \in [n]$, if $\sigma_i, \sigma_i' \in MB(\boldsymbol{\sigma}_{-i})$, we need to prove that $\lambda \sigma_i + (1 - \lambda) \sigma_i' \in MB(\boldsymbol{\sigma}_{-i})$ for any $\lambda \in [0, 1]$ and for any $\boldsymbol{\sigma}_{-i} \in \boldsymbol{\Sigma}_{-i}$.
        Let the supports of $\sigma_i$ and $\sigma_i'$ be defined as $supp(\sigma_i) = \{p_i \in \mathcal{P}_i: \sigma_i(p_i) > 0\}$ and $supp(\sigma_i') = \{p_i \in \mathcal{P}_i: \sigma_i'(p_i) > 0\}$, respectively.
        It is concluded from the definition of the mean-variance best response in Definition \ref{def_best_response_mixed_MV} that $supp(\sigma_i), supp(\sigma_i') \subseteq \underset{p_i \in \mathcal{P}_i}{\argmin} \ \mathrm{Var} \left (\overline{L}^{i}(p_i, \boldsymbol{\sigma}_{-i}) \right ) + \rho \cdot \overline{l}^{i}(p_i, \boldsymbol{\sigma}_{-i})$, which results in $supp(\sigma_i) \cup supp(\sigma_i') \subseteq \underset{p_i \in \mathcal{P}_i}{\argmin} \ \mathrm{Var} \left (\overline{L}^{i}(p_i, \boldsymbol{\sigma}_{-i}) \right ) + \rho \cdot \overline{l}^{i}(p_i, \boldsymbol{\sigma}_{-i})$.
        As a result, using the definition of mean-variance best response, any probability distribution over the set $supp(\sigma_i) \cup supp(\sigma_i')$ is a mean-variance best response to $\boldsymbol{\sigma}_{-i}$.
        The mixed strategy $\lambda \sigma_i + (1 - \lambda) \sigma_i'$ is obviously a valid probability distribution over the set $supp(\sigma_i) \cup supp(\sigma_i')$ for any $\lambda \in [0, 1]$, so $\lambda \sigma_i + (1 - \lambda) \sigma_i' \in MB(\boldsymbol{\sigma}_{-i})$ for any $\lambda \in [0, 1]$ and for any $\boldsymbol{\sigma}_{-i} \in \boldsymbol{\Sigma}_{-i}$ that completes the convexity proof of the set $MB(\boldsymbol{\sigma}_{-i})$.
    \item $\boldsymbol{MB}(\boldsymbol{\sigma})$ has a closed graph:
    $\boldsymbol{MB}(\boldsymbol{\sigma})$ has a closed graph if for any sequence $\{\boldsymbol{\sigma}^m, \widehat{\boldsymbol{\sigma}}^m\} \rightarrow \{\boldsymbol{\sigma}, \widehat{\boldsymbol{\sigma}}\}$ with $\widehat{\boldsymbol{\sigma}}^m \in \boldsymbol{MB}(\boldsymbol{\sigma}^m)$ for all $m \in \mathbb{N}$, we have $\widehat{\boldsymbol{\sigma}} \in \boldsymbol{MB}(\boldsymbol{\sigma})$.
        Proof by contradiction is used to show that $\boldsymbol{MB}(\boldsymbol{\sigma})$ has a closed graph.
        %Proof by contradiction is used to show that $\boldsymbol{RB}(\boldsymbol{\sigma})$ has a closed graph.
        Consider by contradiction that $\boldsymbol{MB}(\boldsymbol{\sigma})$ does not have a closed graph, so there exists a sequence $\{\boldsymbol{\sigma}^m, \widehat{\boldsymbol{\sigma}}^m\} \rightarrow \{\boldsymbol{\sigma}, \widehat{\boldsymbol{\sigma}}\}$ with $\widehat{\boldsymbol{\sigma}}^m \in \boldsymbol{MB}(\boldsymbol{\sigma}^m)$ for all $m \in \mathbb{N}$, but $\widehat{\boldsymbol{\sigma}} \notin \boldsymbol{MB}(\boldsymbol{\sigma})$.
        As a result, there exists some $i \in [n]$ such that $\widehat{\sigma}_i \notin MB(\boldsymbol{\sigma}_{-i})$.
        Using the definition of mean-variance best response in Definition \ref{def_best_response_mixed_MV}, there exists $p_i' \in supp(MB(\boldsymbol{\sigma}_{-i}))$, $\widehat{p}_i \in supp(\widehat{\sigma}_i)$, and some $\epsilon > 0$ such that
        \begin{equation}
            \label{eq_contradiction_1_MV}
            \mathrm{Var} \left (\overline{L}^{i}(p_i', \boldsymbol{\sigma}_{-i}) \right ) + \rho \cdot \overline{l}^{i}(p_i', \boldsymbol{\sigma}_{-i}) < \mathrm{Var} \left (\overline{L}^{i}(\widehat{p}_i, \boldsymbol{\sigma}_{-i}) \right ) + \rho \cdot \overline{l}^{i}(\widehat{p}_i, \boldsymbol{\sigma}_{-i}) - 3 \epsilon.
        \end{equation}
        Since the latencies over edges are continuous random variables and $\boldsymbol{\sigma}_{-i}^m \rightarrow \boldsymbol{\sigma}_{-i}$, for any $\epsilon > 0$, there exists a sufficiently large $m_3$ such that we have the following for $m \geq m_3$:
        \begin{equation}
            \label{eq_contradiction_2_MV}
            \mathrm{Var} \left (\overline{L}^{i}(p_i', \boldsymbol{\sigma}_{-i}^m) \right ) + \rho \cdot \overline{l}^{i}(p_i', \boldsymbol{\sigma}_{-i}^m) < \mathrm{Var} \left (\overline{L}^{i}(p_i', \boldsymbol{\sigma}_{-i}) \right ) + \rho \cdot \overline{l}^{i}(p_i', \boldsymbol{\sigma}_{-i}) + \epsilon.
        \end{equation}
        By adding inequalities with the same direction in Equations \eqref{eq_contradiction_1_MV} and \eqref{eq_contradiction_2_MV}, for $m \geq m_3$ we have 
        \begin{equation}
            \label{eq_contradiction_3_MV}
            \mathrm{Var} \left (\overline{L}^{i}(p_i', \boldsymbol{\sigma}_{-i}^m) \right ) + \rho \cdot \overline{l}^{i}(p_i', \boldsymbol{\sigma}_{-i}^m) < \mathrm{Var} \left (\overline{L}^{i}(\widehat{p}_i, \boldsymbol{\sigma}_{-i}) \right ) + \rho \cdot \overline{l}^{i}(\widehat{p}_i, \boldsymbol{\sigma}_{-i}) - 2 \epsilon.
        \end{equation}
        For the same reason as of Equation \eqref{eq_contradiction_2_MV}, for any $\epsilon > 0$, there exists a sufficiently large $m_4$ such that we have the following for $m \geq m_4$:
        \begin{equation}
            \label{eq_contradiction_4_MV}
            \mathrm{Var} \left (\overline{L}^{i}(\widehat{p}_i, \boldsymbol{\sigma}_{-i}) \right ) + \rho \cdot \overline{l}^{i}(\widehat{p}_i, \boldsymbol{\sigma}_{-i}) < \mathrm{Var} \left (\overline{L}^{i}(\widehat{p}_i^m, \boldsymbol{\sigma}_{-i}^m) \right ) + \rho \cdot \overline{l}^{i}(\widehat{p}_i^m, \boldsymbol{\sigma}_{-i}^m) + \epsilon,
        \end{equation}
        where $\widehat{p}_i^m \in supp(MB(\boldsymbol{\sigma}_{-i}^m))$.
        By adding the inequalities with the same direction in Equations \eqref{eq_contradiction_3_MV} and \eqref{eq_contradiction_4_MV}, for $m \geq \max \{m_3, m_4\}$ we have
        \begin{equation}
            \label{eq_contradiction_5_MV}
            \mathrm{Var} \left (\overline{L}^{i}(p_i', \boldsymbol{\sigma}_{-i}^m) \right ) + \rho \cdot \overline{l}^{i}(p_i', \boldsymbol{\sigma}_{-i}^m) < \mathrm{Var} \left (\overline{L}^{i}(\widehat{p}_i^m, \boldsymbol{\sigma}_{-i}^m) \right ) + \rho \cdot \overline{l}^{i}(\widehat{p}_i^m, \boldsymbol{\sigma}_{-i}^m) - \epsilon.
        \end{equation}
        Equation \eqref{eq_contradiction_5_MV} contradicts the fact that $\widehat{p}_i^m \in supp(MB(\boldsymbol{\sigma}_{-i}^m))$, which completes the proof that $\boldsymbol{MB}(\boldsymbol{\sigma})$ has a closed graph.
\end{enumerate}
As listed above, the mean-variance best response function $\boldsymbol{MB}(\boldsymbol{\sigma})$ satisfies the four conditions of Kakutani's Fixed Point Theorem.
As a direct result, for any finite $n$-player stochastic congestion game, there exists $\boldsymbol{\sigma}^* \in \boldsymbol{\Sigma}$ such that $\boldsymbol{\sigma}^* \in \boldsymbol{MB}(\boldsymbol{\sigma}^*)$, which completes the existence proof of a mean-variance equilibrium for such games.
\Halmos

\section{Proof of Theorem \ref{theorem_existence_CVaR_alpha}}
\label{proof_theorem_CVaR_alpha}
Let $\boldsymbol{CB}: \boldsymbol{\Sigma} \rightarrow \boldsymbol{\Sigma}$ be the CVaR$_\alpha$ best response function where $\boldsymbol{CB}(\boldsymbol{\sigma}) = \big (CB(\boldsymbol{\sigma}_{-1}), CB(\boldsymbol{\sigma}_{-2}),$ $\dots, CB(\boldsymbol{\sigma}_{-N}) \big )$.
It is easy to see that the existence of a fixed point $\boldsymbol{\sigma}^* \in \boldsymbol{\Sigma}$ for the CVaR$_\alpha$ best response function, i.e., $\boldsymbol{\sigma}^* \in \boldsymbol{CB}(\boldsymbol{\sigma}^*)$, proves the existence of a CVaR$_\alpha$ equilibrium.
The following four conditions of the Kakutani’s Fixed Point Theorem are shown to be satisfied for the function $\boldsymbol{CB}(\boldsymbol{\sigma})$ to prove the existence of a fixed point for the function.
\begin{enumerate}[leftmargin=*]
    \item The domain of function $\boldsymbol{CB}(.)$ is a non-empty, compact, and convex subset of a finite dimensional Euclidean space:
    $\boldsymbol{\Sigma}$ is the Cartesian product of non-empty simplices as each player has at least one strategy to play; furthermore, each of the elements of $\boldsymbol{\Sigma}$ is between zero and one, so $\boldsymbol{\Sigma}$ is non-empty, convex, bounded, and closed containing all its limit points.
    %Moreover, if $\boldsymbol{\sigma}_1 \in \boldsymbol{\Sigma}$ and $\boldsymbol{\sigma}_2 \in \boldsymbol{\Sigma}$, 
    \item $\boldsymbol{CB(\boldsymbol{\sigma})} \neq \oldemptyset$, $\forall \boldsymbol{\sigma} \in \boldsymbol{\Sigma}$:
    The set in Equation \eqref{eq_mixed_best_response_CVaR_alpha} is non-empty as minimum exists over a finite number of values.
    As a result, $CB(\boldsymbol{\sigma}_{-i})$ is non-empty for all $i \in [n]$ since it is the set of all probability distributions over the corresponding mentioned non-empty set.
    \item The co-domain of function $\boldsymbol{CB}(.)$ is a convex set for all $\boldsymbol{\sigma} \in \boldsymbol{\Sigma}$:
    It suffices to prove that $CB(\boldsymbol{\sigma}_{-i})$ is a convex set for all $\boldsymbol{\sigma}_{-i} \in \boldsymbol{\Sigma}_{-i}$ and for all $i \in [n]$.
    %Each of the above three cases is proved separately below.
    %$CB(\boldsymbol{\sigma}_{-i})$ is a convex set for all $\boldsymbol{\sigma}_{-i} \in \boldsymbol{\Sigma}_{-i}$:
        For any $i \in [n]$, if $\sigma_i, \sigma_i' \in CB(\boldsymbol{\sigma}_{-i})$, we need to prove that $\lambda \sigma_i + (1 - \lambda) \sigma_i' \in CB(\boldsymbol{\sigma}_{-i})$ for any $\lambda \in [0, 1]$ and for any $\boldsymbol{\sigma}_{-i} \in \boldsymbol{\Sigma}_{-i}$.
        Let the supports of $\sigma_i$ and $\sigma_i'$ be defined as $supp(\sigma_i) = \{p_i \in \mathcal{P}_i: \sigma_i(p_i) > 0\}$ and $supp(\sigma_i') = \{p_i \in \mathcal{P}_i: \sigma_i'(p_i) > 0\}$, respectively.
        It is concluded from the definition of the CVaR$_\alpha$ best response in Definition \ref{def_best_response_mixed_CVaR_alpha} that $supp(\sigma_i), supp(\sigma_i') \subseteq \underset{p_i \in \mathcal{P}_i}{\argmin} \ \mathrm{E} \left [\overline{L}^{i}(p_i, \boldsymbol{\sigma}_{-i}) \Big | \overline{L}^{i}(p_i, \boldsymbol{\sigma}_{-i}) \geq v_\alpha^{i}(p_i, \boldsymbol{\sigma}_{-i}) \right ]$, which results in $supp(\sigma_i) \cup supp(\sigma_i') \subseteq \underset{p_i \in \mathcal{P}_i}{\argmin} \ \mathrm{E} \left [\overline{L}^{i}(p_i, \boldsymbol{\sigma}_{-i}) \Big | \overline{L}^{i}(p_i, \boldsymbol{\sigma}_{-i}) \geq v_\alpha^{i}(p_i, \boldsymbol{\sigma}_{-i}) \right ]$.
        As a result, using the definition of CVaR$_\alpha$ best response, any probability distribution over the set $supp(\sigma_i) \cup supp(\sigma_i')$ is a CVaR$_\alpha$ best response to $\boldsymbol{\sigma}_{-i}$.
        The mixed strategy $\lambda \sigma_i + (1 - \lambda) \sigma_i'$ is obviously a valid probability distribution over the set $supp(\sigma_i) \cup supp(\sigma_i')$ for any $\lambda \in [0, 1]$, so $\lambda \sigma_i + (1 - \lambda) \sigma_i' \in CB(\boldsymbol{\sigma}_{-i})$ for any $\lambda \in [0, 1]$ and for any $\boldsymbol{\sigma}_{-i} \in \boldsymbol{\Sigma}_{-i}$ that completes the convexity proof of the set $CB(\boldsymbol{\sigma}_{-i})$.
    \item $\boldsymbol{CB}(\boldsymbol{\sigma})$ has a closed graph:
    $\boldsymbol{CB}(\boldsymbol{\sigma})$ has a closed graph if for any sequence $\{\boldsymbol{\sigma}^m, \widehat{\boldsymbol{\sigma}}^m\} \rightarrow \{\boldsymbol{\sigma}, \widehat{\boldsymbol{\sigma}}\}$ with $\widehat{\boldsymbol{\sigma}}^m \in \boldsymbol{CB}(\boldsymbol{\sigma}^m)$ for all $m \in \mathbb{N}$, we have $\widehat{\boldsymbol{\sigma}} \in \boldsymbol{CB}(\boldsymbol{\sigma})$.
        Proof by contradiction is used to show that $\boldsymbol{CB}(\boldsymbol{\sigma})$ has a closed graph.
        %Proof by contradiction is used to show that $\boldsymbol{RB}(\boldsymbol{\sigma})$ has a closed graph.
        Consider by contradiction that $\boldsymbol{CB}(\boldsymbol{\sigma})$ does not have a closed graph, so there exists a sequence $\{\boldsymbol{\sigma}^m, \widehat{\boldsymbol{\sigma}}^m\} \rightarrow \{\boldsymbol{\sigma}, \widehat{\boldsymbol{\sigma}}\}$ with $\widehat{\boldsymbol{\sigma}}^m \in \boldsymbol{CB}(\boldsymbol{\sigma}^m)$ for all $m \in \mathbb{N}$, but $\widehat{\boldsymbol{\sigma}} \notin \boldsymbol{CB}(\boldsymbol{\sigma})$.
        As a result, there exists some $i \in [n]$ such that $\widehat{\sigma}_i \notin CB(\boldsymbol{\sigma}_{-i})$.
        Using the definition of CVaR$_\alpha$ best response in Definition \ref{def_best_response_mixed_CVaR_alpha}, there exists $p_i' \in supp(CB(\boldsymbol{\sigma}_{-i}))$, $\widehat{p}_i \in supp(\widehat{\sigma}_i)$, and some $\epsilon > 0$ such that
        \begin{equation}
            \label{eq_contradiction_1_CVaR_alpha}
            \mathrm{E} \left [\overline{L}^{i}(p_i', \boldsymbol{\sigma}_{-i}) \Big | \overline{L}^{i}(p_i', \boldsymbol{\sigma}_{-i}) \geq v_\alpha^{i}(p_i', \boldsymbol{\sigma}_{-i}) \right ] < \mathrm{E} \left [\overline{L}^{i}(\widehat{p}_i, \boldsymbol{\sigma}_{-i}) \Big | \overline{L}^{i}(\widehat{p}_i, \boldsymbol{\sigma}_{-i}) \geq v_\alpha^{i}(\widehat{p}_i, \boldsymbol{\sigma}_{-i}) \right ] - 3 \epsilon.
        \end{equation}
        Since the latencies over edges are continuous random variables and $\boldsymbol{\sigma}_{-i}^m \rightarrow \boldsymbol{\sigma}_{-i}$, for any $\epsilon > 0$, there exists a sufficiently large $m_5$ such that we have the following for $m \geq m_5$:
        \begin{equation}
            \label{eq_contradiction_2_CVaR_alpha}
            \mathrm{E} \left [\overline{L}^{i}(p_i', \boldsymbol{\sigma}_{-i}^m) \Big | \overline{L}^{i}(p_i', \boldsymbol{\sigma}_{-i}^m) \geq v_\alpha^{i}(p_i', \boldsymbol{\sigma}_{-i}^m) \right ] < \mathrm{E} \left [\overline{L}^{i}(p_i', \boldsymbol{\sigma}_{-i}) \Big | \overline{L}^{i}(p_i', \boldsymbol{\sigma}_{-i}) \geq v_\alpha^{i}(p_i', \boldsymbol{\sigma}_{-i}) \right ] + \epsilon.
        \end{equation}
        By adding inequalities with the same direction in Equations \eqref{eq_contradiction_1_CVaR_alpha} and \eqref{eq_contradiction_2_CVaR_alpha}, for $m \geq m_5$ we have 
        \begin{equation}
            \label{eq_contradiction_3_CVaR_alpha}
            \mathrm{E} \left [\overline{L}^{i}(p_i', \boldsymbol{\sigma}_{-i}^m) \Big | \overline{L}^{i}(p_i', \boldsymbol{\sigma}_{-i}^m) \geq v_\alpha^{i}(p_i', \boldsymbol{\sigma}_{-i}^m) \right ] < \mathrm{E} \left [\overline{L}^{i}(\widehat{p}_i, \boldsymbol{\sigma}_{-i}) \Big | \overline{L}^{i}(\widehat{p}_i, \boldsymbol{\sigma}_{-i}) \geq v_\alpha^{i}(\widehat{p}_i, \boldsymbol{\sigma}_{-i}) \right ] - 2 \epsilon.
        \end{equation}
        For the same reason as of Equation \eqref{eq_contradiction_2_CVaR_alpha}, for any $\epsilon > 0$, there exists a sufficiently large $m_6$ such that we have the following for $m \geq m_6$:
        \begin{equation}
            \label{eq_contradiction_4_CVaR_alpha}
            \mathrm{E} \left [\overline{L}^{i}(\widehat{p}_i, \boldsymbol{\sigma}_{-i}) \Big | \overline{L}^{i}(\widehat{p}_i, \boldsymbol{\sigma}_{-i}) \geq v_\alpha^{i}(\widehat{p}_i, \boldsymbol{\sigma}_{-i}) \right ] < \mathrm{E} \left [\overline{L}^{i}(\widehat{p}_i^m, \boldsymbol{\sigma}_{-i}^m) \Big | \overline{L}^{i}(\widehat{p}_i^m, \boldsymbol{\sigma}_{-i}^m) \geq v_\alpha^{i}(\widehat{p}_i^m, \boldsymbol{\sigma}_{-i}^m) \right ] + \epsilon,
        \end{equation}
        where $\widehat{p}_i^m \in supp(CB(\boldsymbol{\sigma}_{-i}^m))$.
        By adding the inequalities with the same direction in Equations \eqref{eq_contradiction_3_CVaR_alpha} and \eqref{eq_contradiction_4_CVaR_alpha}, for $m \geq \max \{m_5, m_6\}$ we have
        \begin{equation}
            \label{eq_contradiction_5_CVaR_alpha}
            \mathrm{E} \left [\overline{L}^{i}(p_i', \boldsymbol{\sigma}_{-i}^m) \Big | \overline{L}^{i}(p_i', \boldsymbol{\sigma}_{-i}^m) \geq v_\alpha^{i}(p_i', \boldsymbol{\sigma}_{-i}^m) \right ] < \mathrm{E} \left [\overline{L}^{i}(\widehat{p}_i^m, \boldsymbol{\sigma}_{-i}^m) \Big | \overline{L}^{i}(\widehat{p}_i^m, \boldsymbol{\sigma}_{-i}^m) \geq v_\alpha^{i}(\widehat{p}_i^m, \boldsymbol{\sigma}_{-i}^m) \right ] - \epsilon.
        \end{equation}
        Equation \eqref{eq_contradiction_5_CVaR_alpha} contradicts the fact that $\widehat{p}_i^m \in supp(CB(\boldsymbol{\sigma}_{-i}^m))$, which completes the proof that $\boldsymbol{CB}(\boldsymbol{\sigma})$ has a closed graph.
\end{enumerate}
As listed above, the CVaR$_\alpha$ best response function $\boldsymbol{CB}(\boldsymbol{\sigma})$ satisfies the four conditions of Kakutani's Fixed Point Theorem.
As a direct result, for any finite $n$-player stochastic congestion game, there exists $\boldsymbol{\sigma}^* \in \boldsymbol{\Sigma}$ such that $\boldsymbol{\sigma}^* \in \boldsymbol{CB}(\boldsymbol{\sigma}^*)$, which completes the existence proof of a CVaR$_\alpha$ equilibrium for such games.
\Halmos

\end{appendices}

\end{document}